\documentclass[12pt]{article}

\usepackage{amsmath,amssymb,graphicx,multirow,xspace,slashed}
\usepackage[colorlinks=true,urlcolor=blue,anchorcolor=blue,citecolor=blue,filecolor=blue,linkcolor=blue,menucolor=blue,pagecolor=blue]{hyperref}
\usepackage[compress,numbers]{natbib}
\usepackage{subfigure, placeins}
\usepackage{booktabs}
\usepackage{blindtext, rotating}
\usepackage{afterpage}
\usepackage{enumitem}
\usepackage{ marvosym }
\usepackage{authblk}
\usepackage{verbatim}
\usepackage{soul} 

\usepackage{floatrow}
\newfloatcommand{capbtabbox}{table}[][\FBwidth]

\usepackage[font=footnotesize,labelfont=bf]{caption}

\usepackage{lineno}

\allowdisplaybreaks
\long\def\symbolfootnote[#1]#2{\begingroup%
\def\thefootnote{\fnsymbol{footnote}}\footnote[#1]{#2}\endgroup}

\newcommand{\newc}{\newcommand}
\newc{\gsim}{\lower.7ex\hbox{$\;\stackrel{\textstyle>}{\sim}\;$}}
\newc{\lsim}{\lower.7ex\hbox{$\;\stackrel{\textstyle<}{\sim}\;$}}
\newc{\gev}{\,{\rm GeV}}
\newc{\mev}{\,{\rm MeV}}
\newc{\ev}{\,{\rm eV}}
\newc{\kev}{\,{\rm keV}}
\newc{\tev}{\,{\rm TeV}}

\def\Im{\mathop{\rm Im}}
\def\Re{\mathop{\rm Re}}

\newc{\mz}{M_Z}
\newc{\mpl}{M_*}
\newc{\mw}{m_{\rm weak}}
\newc{\nr}[1]{N^c_R{}_{#1}}

\newcommand{\s}{\sigma}

\renewcommand{\dag}{\dagger}

\renewcommand{\O}{O}

\def\beq{\begin{equation}}
\def\eeq{\end{equation}}
\newcommand{\bea}{\begin{eqnarray}\begin{aligned}}
\newcommand{\eea}{\end{aligned}\end{eqnarray}}
\def\bitem{\begin{itemize}}
\def\eitem{\end{itemize}}
\newc{\ie}{{\it i.e.}}          \newc{\etal}{{\it et al.}}
\newc{\eg}{{\it e.g.}}          \newc{\etc}{{\it etc.}}
\newc{\cf}{{\it c.f.}}

\newcommand{\CO}{O}

 \numberwithin{equation}{section}

\newcommand\fverb{\setbox\fverbbox=\hbox\bgroup\verb}
\newcommand\fverbdo{\egroup\medskip\noindent%
            \fbox{\unhbox\fverbbox}\ }
\newcommand\fverbit{\egroup\item[\fbox{\unhbox\fverbbox}]}
\newbox\fverbbox

\usepackage[usenames,dvipsnames]{xcolor}

\begin{document}

\baselineskip 0.6cm

\begin{titlepage}

\thispagestyle{empty}

\begin{center}

\vskip 1cm

{\Large \bf Dark Matter and the Higgs in Natural SUSY}

\vskip 1.0cm
{\large Aria Basirnia, Sebastian Macaluso and David Shih }
\vskip 0.4cm
{\it NHETC, Department of Physics and Astronomy, Rutgers University, \\ Piscataway, NJ 08854} \\
\vskip 2.0cm

\end{center}

\begin{abstract}

\noindent Null results from dark matter (DM) direct detection experiments and the 125 GeV Higgs both pose serious challenges to minimal supersymmetry. In this paper, we propose a simple extension of the MSSM that economically solves both problems: a ``dark sector" consisting of a singlet and a pair of $SU(2)$ doublets. Loops of the dark sector fields help lift the Higgs mass to 125 GeV consistent with naturalness, while the lightest fermion in the dark sector can be viable thermal relic DM, provided that it is mostly singlet. The DM relic abundance is controlled by $s$-wave annihilation to tops and Higgsinos, leading to a tight relation between the relic abundance and the spin-dependent direct detection cross section. As a result, the model will be fully probed by the next generation of direct detection experiments. Finally we discuss the discovery potential at LHC Run II.

\end{abstract}

\end{titlepage}

\setcounter{page}{1}

\section{Introduction and Summary}

The MSSM paradigm is under siege from both the LHC and dark matter (DM) direct detection. 
The Higgs mass at tree-level in the MSSM is famously bounded by $m_Z$, and relying on radiative corrections from stops and other particles in the MSSM forces the stops to be either at least $\sim 10$~TeV or their $A$-terms to be multi-TeV (for recent reviews and original references, see e.g.\   \cite{Martin:1997ns,Halkiadakis:2014qda,Draper:2016pys}). 
Together with the null direct search results at the LHC, this puts the fine-tuning in the MSSM at the percent level or worse. 
Meanwhile, to evade stringent direct and indirect detection bounds,  thermal relic neutralino DM in the MSSM must rely on increasingly contrived numerical accidents (well-tempering, blind spots, funnels, co-annihilations) or an increasingly heavy SUSY scale (e.g.\ $\sim 1$~TeV Higgsinos or $\sim 2-3$~TeV winos) (see e.g.\ \cite{Cheung:2012qy, Han:2013gba, Crivellin:2015bva} for recent comprehensive studies). The latter constitutes a DM version of the little hierarchy problem, whereby the WIMP miracle's preference for TeV-scale DM (as opposed to 100~GeV scale DM) is in tension with naturalness.

This strongly motivates looking beyond the MSSM for both the source of the Higgs mass and dark matter. Although it is logically possible that different sectors are independently responsible for the Higgs mass and dark matter,  it is interesting to contemplate more elegant and economical models where a single sector generates both. In this paper, we will study such a model. We will show how to achieve a 125 GeV Higgs and thermal relic WIMP DM consistent with all existing constraints, while greatly ameliorating the fine-tuning, by just adding a pair of $SU(2)$ doublets $L$, $\bar L$ and a singlet $S$ to the MSSM. With a $Z_2$ ``DM parity" that keeps the lightest state in the dark sector stable, together with matter parity from the MSSM, the most general renormalizable superpotential for this ``dark sector" is: 
\bea\label{eq:superpotentialintro}
W	&= {1\over2}M_S S^2 + M_L L \bar{L} + k_u H_u L S - k_d H_d \bar{L} S \\
\eea
Although it would be interesting to also consider phases, we will focus on real couplings in this paper for simplicity. Then without loss of further generality, we can take $M_S$ and $M_L$ to be positive.

The idea of extending the Standard Model (SM) with a ``singlet-doublet DM" sector has been studied previously in \cite{Mahbubani:2005pt, Hisano:2010bx, Cohen:2011ec, Cheung:2013dua, Abe:2014gua, Calibbi:2015nha, Freitas:2015hsa,Banerjee:2016hsk}, motivated by minimality and by the fact that it is a simple generalization of the well-studied bino/Higgsino system of the MSSM. The idea of lifting the Higgs mass with loops of vector-like matter has also been well-studied \cite{Moroi:1991mg,Moroi:1992zk, Babu:2004xg, Babu:2008ge, Martin:2009bg,  Graham:2009gy, Martin:2010dc, Martin:2010kk, Asano:2011zt, Endo:2011mc, Evans:2011uq, Moroi:2011aa,Endo:2011xq,Endo:2012rd,Martin:2012dg,Endo:2012cc,Iwamoto:2013kla, Nakai:2015swg}. But to our knowledge, the two ideas have never been combined before.\footnote{A related idea \cite{Abdullah:2015zta,Abdullah:2016avr} is to use vector-like matter to boost the Higgs mass while simultaneously opening up new annihilation channels for bino-like DM.} Combining these two ideas leads to some important differences with previous works. 

First, unlike in previous works on lifting the Higgs mass, our dark sector cannot be truly vector-like. The scalar soft mass-squareds of the dark sector must be positive in order to lift the Higgs mass,  making our DM the lightest fermion in the dark sector. It cannot be a Dirac fermion, otherwise it would be ruled out by many orders of magnitude by $Z$-mediated spin-independent (SI) direct detection. Instead, we make the dark sector fermions Majorana (as shown in (\ref{eq:superpotentialintro})) by having only one singlet and not a vector-like pair of them. This only has a minor effect on the contribution to the Higgs mass in this model, which we fully take into account.  We will find that a $m_h=125$~GeV Higgs can be achieved with the fine-tuning coming from the DM being only $\sim$10\%, provided that $k_u\sim\CO(1)$. 

Second, we differ from the singlet-doublet DM models in that we are supersymmetrizing everything.\footnote{Actually, in \cite{Hisano:2010bx} they also added singlets and doublets to the MSSM. However, they considered soft masses purely from GMSB (whereas we are agnostic) and therefore they never have mostly-singlet fermionic DM. Moreover they fix $k_u = k_d = 0.3$ whereas we have them as free parameters. Finally, they do not calculate the contribution to the Higgs mass from the dark sector.} A priori, the parameter space of the entire model (MSSM+dark sector)  is vast, but most of the soft parameters do not play a significant role in the analysis. As seen in (\ref{eq:superpotentialintro}), our dark sector only couples directly to the Higgs sector and the EW gauge sector of the MSSM. We will keep the Higgsinos light ($\lesssim 300$~GeV), since they contribute to the  fine-tuning of the EW scale at tree level. As a result, DM annihilation to light Higgsinos through superpartners in the dark sector plays a major role in determining the relic abundance of the DM. Meanwhile, it does not change our analysis qualitatively to decouple all other MSSM superpartners (effectively at the $\sim$ TeV scale). This is further motivated by the null results from the LHC.

\begin{figure}[t]
\centering
\includegraphics[scale=0.2]{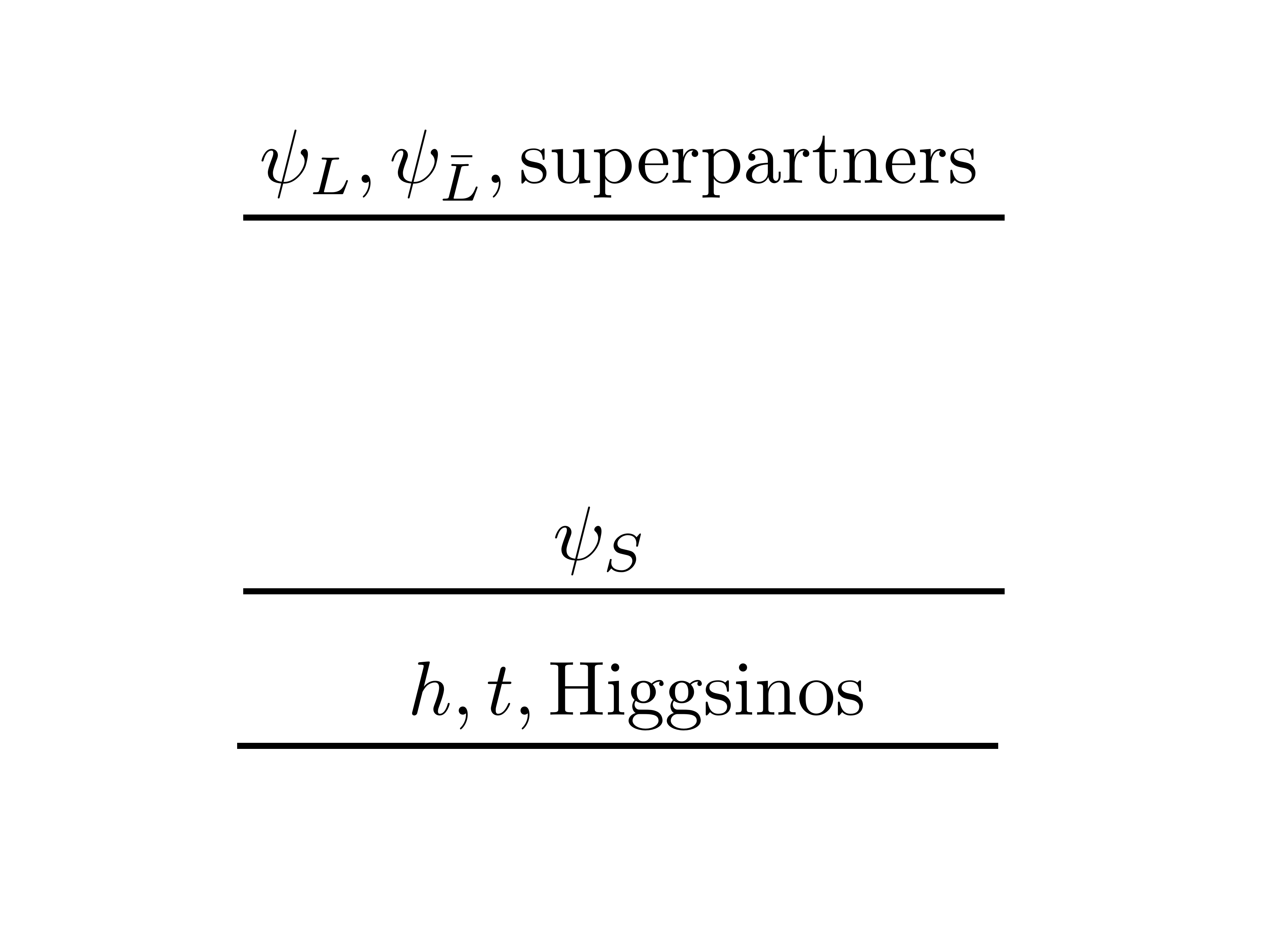}
\caption{\small{A typical, viable spectrum of the model. $\psi_S$, $\psi_L$, $\psi_{\bar L}$ are the fermionic components of the dark sector fields. Superpartners include scalar components of the dark sector and superparticles in MSSM.}}
\label{fig:spectrum}
\end{figure}

We will further simplify the analysis of the model by focusing on the regime where the dark matter $\chi$ is mostly-singlet, i.e.\ $M_S<M_L$ and $v\ll M_L, \, M_L-M_S$. As we will argue in much more detail in section \ref{sec:mostly_singlet}, this regime is absolutely necessary in order to evade direct detection bounds while raising the Higgs mass without fine-tuning.  A key part of the argument, which distinguishes this from the bino/Higgsino system in the MSSM, is that $k_u$ must be $\CO(1)$ in order to lift the Higgs mass without fine-tuning. This eliminates both the well-tempered regime and the mostly-doublet regime vis a vis DM direct detection. The mostly-doublet regime is further unpromising because (by analogy with pure Higgsino DM in the MSSM) it would require a DM mass in excess of 1~TeV, and this would greatly exacerbate the fine-tuning problem,  since the rest of the dark sector would have to be even heavier. This leaves the mostly-singlet regime, where the analysis of the model greatly simplifies, and we are able to understand all the features of the model with simple analytic formulas. A cartoon spectrum of the model that describes these hierarchies qualitatively is shown in fig.~\ref{fig:spectrum}.

In this work, we will assume the simplest DM scenario, namely that $\chi$ is a thermal relic comprising all of the DM. In the mostly-singlet limit with $k_u\sim 1$, we will show that the thermal relic abundance is controlled by just two DM annihilation channels:  $s$-wave $t\bar t$ (through $s$-channel $Z$ exchange) and  $s$-wave Higgsinos (through $t$-channel superpartner exchange). Assuming $M_S\ll M_L$ for simplicity, we find:
\beq
\sigma v_\chi \approx {3k_u^4 m_t^2\over 32\pi M_L^4} + {(k_u^2+k_d^2)^2\mu^2\over 16\pi (M_L^2+m^2)^2}\label{sigmavintro}
\eeq
where $m$ is a common soft mass for the dark sector scalars. As noted above, the second term coming from Higgsinos is a major difference from the non-supersymmetric singlet-doublet DM models that have been studied previously. Having more annihilation channels increases $\sigma v_\chi$, making it possible to have smaller effective couplings between the DM and the SM. This opens up more parameter space  that is not ruled out by direct detection experiments and yet still has the correct thermal relic abundance, as compared to the non-SUSY singlet-doublet models.

Interestingly, the DM mass drops out of the annihilation cross section (\ref{sigmavintro}) in the mostly-singlet limit. 
The WIMP miracle becomes one for the mediator scale, not the WIMP mass!  With $k_u\sim k_d\sim 1$, $m\sim M_L$ and $\mu\lesssim 300$~GeV, mediator scales of $M_L\sim 1-2$~TeV are implied by the thermal relic constraint. Meanwhile the DM can be much lighter than this, alleviating the DM little hierarchy problem. It is also interesting to contrast this with the mostly bino limit of the bino/Higgsino system in the MSSM. There the annihilation cross section is not large enough, being suppressed by $g_1^4$ instead of $k_u^4$. Our model (and singlet-doublet DM more generally) gets around this in the mostly-singlet regime with $\CO(1)$ Yukawa couplings that are free parameters, not fixed to be $g_1$ by supersymmetry. 

Meanwhile, DM direct detection in these models is completely controlled by the effective couplings of the DM to the Higgs and $Z$ respectively:
\bea\label{eq:chcZL}
\delta{\mathcal L} = c_h h \bar\psi_\chi\psi_\chi + c_Z Z_\mu \bar\psi_\chi \gamma^\mu\gamma^5 \psi_\chi
\eea
As is well-known, $c_h$ ($c_Z$) controls the SI (SD) direct detection cross section. For direct detection, as we will review, the current best bounds for our DM mass range of interest ($100\lesssim m_{DM}\lesssim 1000$~GeV) come from LUX \cite{Akerib:2016vxi} and IceCube \cite{Aartsen:2016exj}. We will convert the official experimental results, which are phrased in terms of the DM-nucleon cross section, into limits on $c_h$ and $c_Z$. Furthermore, in the mostly-singlet limit, we will obtain simple analytic expressions for $c_h$ and $c_Z$. We will see that $c_h$ can be naturally small enough for mostly-singlet DM, due to suppression from the heavier doublet scale, as well as a mild blind-spot cancellation:
\beq
c_h \approx -{m_\chi+ {2k_d M_L\over k_u \tan\beta}\over\sqrt{2}v} {k_u^2v^2\over M_L^2} 
\eeq
provided that $k_u\sim k_d$. We should emphasize here that the Higgs mass depends not just on $c_h$ but also on the effective Yukawa couplings between the Higgs and the other dark sector particles. So even dialing $c_h\to 0$ does not qualitatively affect the Higgs mass calculation. 
Meanwhile $c_Z$ is given in the mostly-singlet limit by:
\beq
c_Z\approx -{g_2\over 4c_W}{k_u^2v^2\over M_L^2}
\eeq
According to our discussion above, after fixing the thermal relic density constraint $\Omega_{DM}^{obs}h^2\approx 0.12$, $c_Z$ is essentially fixed to lie within a narrow range which depends primarily on the Higgsino mass $\mu$. Therefore imposing the relic density constraint essentially fixes the SD cross section. Fortunately, this value is not ruled out yet, but the next generation of DM experiments (e.g.\ Xenon1T \cite{Aprile:2015uzo}, LZ \cite{Akerib:2015cja}) should completely rule out or discover this model.

Although direct detection is controlled by $c_Z$ and $c_h$, the other facets of the model (relic abundance, Higgs mass) depend on more than just these couplings, so our model does not fit completely into the framework of $Z$- and $h$-portal DM. For instance, we mentioned above that the Higgsino cross section arises entirely from $t$-channel superpartner exchange. Also, we find that DM annihilation to dibosons is suppressed more than would be the case in $Z$ and $h$ portal models, in part due to $t$-channel exchange of doublet fermions. Similar comments apply to the effective operator formalism: our DM is generally not light enough compared to the mediator scale (the doublet mass) for the annihilation to be accurately captured by effective operators. Evidently, the complete model (\ref{eq:superpotentialintro}) is required for an accurate analysis. This illustrates the shortcomings and limitations of both simplified models and effective operator approaches to dark matter. 

We have focused primarily on the standard direct detection searches in this work, because other indirect probes of our dark sector are far less sensitive. For example, the Fermi experiment and others have searched for energetic photons produced through DM annihilating at the centers of dwarf galaxies. For DM masses above $100$ GeV, Fermi does not constrain any point with the right relic-abundance \cite{Ackermann:2015zua}, assuming (as is the case for us) that the relic abundance is determined by $s$-wave annihilation. Meanwhile, searches at colliders and electroweak precision tests (EWPT) could have put constraints on our model. However as we will discuss further in section \ref{sec:LHC Phenomenology}, LHC bounds  {\cite{Khachatryan:2014rra, Aad:2015zva, Chatrchyan:2012tea, Aad:2012fw}} on $c_h$ and $c_Z$ from monojets+MET and monophoton+MET are orders of magnitude weaker than direct detection for the range of DM masses that we are interested in. We will briefly discuss mono-($W$,$Z$,$h$)+MET and show how it could probe the low end of DM masses ($m_{DM}\sim 200$~GeV) in our model, with 300/fb at LHC Run II. Finally,  limits from Higgs and $Z$ invisible width do not apply to the mass range of DM that we consider in this work, and we checked that contributions to the $S$ and $T$ parameters are well within the acceptable range, in agreement with previous studies of these variables in closely-related models \cite{Abe:2014gua,Martin:2009bg}.

In this paper, we will analyze the model using a combination of simple, approximate analytic expressions valid in the mostly-singlet regime, and more precise numerical methods that take into account the full suite of one (and even two) loop threshold corrections. The analytic approach, while being reasonably accurate, is primarily a source of intuition and understanding. The numerical approach is meant to be more accurate and to provide us with the quantitative results. Clearly, having both numerics and analytics is a vital source of cross-checks, giving us greater confidence in our results.  

Our numerical methods are based on publicly available codes. Our starting point was the powerful SARAH 4.5.8  framework \cite{Staub:2013tta} for automated analysis of general models. Once we properly defined our model, SARAH automatically generated source code for SPheno 3.3.7 \cite{Porod:2003um,Porod:2011nf} and for micrOMEGAs 4.1.8 \cite{Belanger:2014vza}.\footnote{We are extremely grateful to Florian Staub for his time and patience in helping us set up the SARAH model and link it to these other codes.} The former calculates the spectrum while the latter calculates the DM relic abundance and direct detection cross sections. In our numerical calculations, all MSSM soft masses as well as gauginos are taken to be at 1 TeV, and the $A$-terms are set to zero.  As noted earlier, since $\mu$ appears at tree level in fine-tuning of the electroweak scale we treat it differently. We pick $\mu = 300 \,\text{GeV}$  in our numerical calculations which corresponds roughly to $10\%$ fine-tuning. We also consider $\mu = 100 \, \text{GeV}$ to see the effect of $\mu$ on our analysis. Finally, to saturate the tree level contribution to the Higgs mass, we take the other Higgses to be heavy and in the decoupling limit, and we take $\tan\beta=10$.

The outline of our paper is as follows. In section \ref{sec:model} we introduce the model. Then in section \ref{sec:DMDD}, we derive direct detection limits from LUX and IceCube on the effective couplings $c_h$ and $c_Z$. We will emphasize that these results are general and are not limited to the model we consider in this work. In section \ref{Higgs mass and Fine Tuning} we compute the one-loop corrections to the Higgs mass from the new particles in our model, and we discuss fine-tuning.  We argue in section \ref{sec:mostly_singlet} that the mostly singlet case is the only viable scenario. In the mostly singlet limit, we provide analytic expressions for dark matter annihilation in the early universe for our model in section \ref{DM annihilation}. In section \ref{sec:results} we put everything together to show the viable parameter space that satisfies all direct detection constraints while having the right relic abundance and Higgs mass. Here we demonstrate quantitatively that requiring $\chi$ to be all of the DM essentially fixes $c_Z$ (and hence $\sigma^{SD}$) to a unique value which is not yet ruled out by direct detection, but will be fully within reach of the next generation of experiments. We conclude by studying the collider signatures for LHC Run II and the UV behavior of the model, and  giving suggestions on future directions on section \ref{sec:Outlook}. Technical details and validations are reserved for three appendices. In appendix \ref{appendix:DD_to_ch/cZ} we review the derivation of the direct detection cross sections from effective DM nucleon couplings. We validate our numerical and analytical calculations of the Higgs mass in appendix \ref{appendix:Validating_Higgs_mass}. Finally we provide analytical cross sections for DM production at LHC II in appendix \ref{appendix:LHC analytics}.


\section{The Model}
\label{sec:model}

We begin by describing the model in more detail. We add to the MSSM a ``dark sector" consisting of a vector-like pair of $SU(2)$ doublets $L$, $\bar L$ and a gauge singlet $S$.\footnote{To keep gauge coupling unification as in MSSM, we can assume $L$ and $\bar{L}$ are part of complete $\mathbf{5}$ and $\bar{\mathbf{5}}$ multiplets of $SU(5)$. We take their colored partners to be heavy and decoupled for simplicity. 
} The dark sector is equipped with an unbroken $Z_2^{DM}$ parity symmetry under which all new fields are odd and all MSSM fields are even. This makes the lightest new state stable and a DM candidate.  Finally, we assume MSSM matter parity, under which all the dark sector fields have the same charge; otherwise there will be additional, potentially dangerous terms.\footnote{The assumption of matter parity implies another stable particle -- either the LSP in the MSSM, or the gravitino. Either way, we assume the parameters are such that this will add a negligible additional component to the thermal relic density. This would be the case, for instance, if the LSP is a light Higgsino.} The transformation properties of the dark sector under the gauge and global symmetries is summarized in tab.~\ref{tab:matter content}.

The most generic superpotential consistent with these symmetries is: 
\bea\label{eq:superpotential}
\delta W	&= {1\over2}M_S S^2 + M_L L \bar{L} + k_u H_u L S - k_d H_d \bar{L}S \\
\eea
The superpotential has four new parameters in addition to the MSSM: $M_L, M_S, k_u, k_d$. There is one physical complex phase, but as discussed in the introduction, we will take these parameters to be real in this paper. In this case, there is still a physical sign. We will take $M_L$, $M_S$ and $k_u$ to be positive and put the sign into $k_d$.

\begin{table}[t]
$$
\begin{array}{|c|c|c|c|c|c|}\hline
 & SU(3)_c & SU(2)_L & U(1)_Y  & Z_2^{DM} & Z_2^M \\\hline
 L 
 & {\bf 1} & {\bf 2} & -{1\over 2} & -1 & 1  \\
 \bar{L} 
 & {\bf 1} & {\bf 2} & {1\over 2} & -1 & 1 \\
 S & {\bf 1} & {\bf 1} & 0 & -1 & 1 \\
 \hline
\end{array}
$$
\caption{Gauge and global symmetries of the dark sector.}
\label{tab:matter content}
\end{table}

For the soft SUSY-breaking Lagrangian,  for simplicity we take the minimal case with equal soft mass-squareds and no $A$- or $B$-terms:
\bea\label{eq:msoft}
\delta\mathcal{L}_{\text{soft}} = -m^2 (|\bar{\ell}|^2 + |\ell|^2 + |s|^2) 
\eea
(We denote the scalar components of the dark sector superfields with lowercase letters.) Allowing different soft masses for the different fields will not change most of the discussion in this paper, only the contributions to Higgs mass.

As we want this new sector to increase the lightest Higgs mass analogous to the MSSM stops, we assume that $m^2>0$. This implies that the DM candidate is a fermion. Furthermore it is Majorana, thanks to the fact that we have included only one singlet in the theory. Had we started with a Dirac pair of $S$ and $\bar{S}$ and defined the mass term as $M_S S \bar{S}$, our dark matter would have had a vector-like coupling to the $Z$. In that case it would have been impossible to hide it from SI direct detection experiments while keeping the interesting features of our model.

After EWSB, neutral fields in the dark sector mix through the Yukawa couplings in (\ref{eq:superpotential}). The fermion mass matrix of the neutral states is:
\beq
\label{eq:mass matrix}
{\mathcal M} = \left(\begin{matrix} M_S & \hat k_u v  & \hat k_d v \\ \hat k_u v   & 0 & M_L \\ \hat k_d v   & M_L & 0 \end{matrix} \right),
\eeq
where we have introduced $\hat{k}_u \equiv k_u \sin\beta$ and $\hat{k}_d \equiv k_d \cos\beta$, with $\tan\beta=v_u/v_d$ and   $v_u^2+v_d^2=v^2\approx(174 \, \text{GeV})^2$ as usual. We take large $\text{tan}\beta = 10$ in this paper to saturate the upper bound on the tree level Higgs mass.
The mass matrix is diagonalized by $U {\mathcal M} U^\dag= {\mathcal M}_{\text{diag}}$.
The spectrum of the model consists of three Majorana fermions with masses $m_{\chi_1}< m_{\chi_2}<m_{\chi_3}$ and a Dirac charged fermion with mass $m_{\chi^\pm} = M_L$. The dark matter candidate is then $\chi\equiv \chi_1$.

We note that the fermionic part of our dark sector is analogous to Bino-Higgsino DM in the MSSM (with everything else decoupled), except that  in the Bino-Higgsino system, we effectively have $k_u = k_d = g'/\sqrt{2}$, whereas here $k_u$ and $k_d$ are general. In fact, as discussed in the introduction, here we will be primarily interested in $k_u$, $k_d\sim \CO(1)$.

After rotating to the mass eigenbasis, DM-$Z$ and DM-Higgs couplings are generated:
\bea\label{eq:chcZL}
\delta{\mathcal L} = c_h h \bar\psi_\chi\psi_\chi + c_Z Z_\mu \bar\psi_\chi \gamma^\mu\gamma^5 \psi_\chi
\eea
where $\psi_\chi = ( \chi,\chi^\dagger)^T$ is a 4-component Majorana fermion and $c_h$ and $c_Z$ are given by:
\bea\label{eq:chdef}
c_{h }&=\frac{1}{\sqrt{2}}{\rm Re}(\hat k_u  U^*_{11} U^*_{12}+\hat k_d  U^*_{11} U^*_{13})\\
&=\frac{v}{\sqrt{2}}\left(\frac{m_{\chi}(\hat{k}_d^2+\hat{k}_u^2)+2\hat{k}_d\hat{k}_u M_L}{M_L^2+ 2 M_S m_{\chi}-3 m^2_{\chi}+v^2(\hat{k}_d^2 + \hat{k}_u^2)}\right)
\eea
and
\bea\label{eq:cZdef}
c_{Z }&=\frac{g_2}{4 c_W}\left(|U_{12}|^2-|U_{13}|^2\right)\\
&=\frac{g_2}{4c_W} \frac{(M_L^2-m_{\chi}^2) v^2(\hat{k}_d^2-\hat{k}_u^2)}{  (M_L^2-m_{\chi}^2)^2+v^2((\hat{k}_u^2+\hat{k}_d^2)(M_L^2+m_{\chi}^2) + 4\hat{k}_d \hat{k}_u M_L m_{\chi})}
\eea
As is well-known \cite{Cohen:2011ec, Cheung:2013dua,Abe:2014gua, Calibbi:2015nha, Freitas:2015hsa}, $c_h$ and $c_Z$ play an important role in the analysis of singlet-doublet DM:  they are entirely responsible for SI and SD direct detection, respectively. In the next section, we will review the current direct detection constraints on $c_h$ and $c_Z$.


\section{DM Direct Detection through the $h$ and $Z$ Portals}
\label{sec:DMDD}

In the DM mass range of interest (100~GeV $\lesssim m_{DM}\lesssim$ 1~TeV), the LUX experiment currently sets the best bound on SI elastic WIMP-nucleon scattering \cite{ Akerib:2016vxi}. Meanwhile, the best limits for SD elastic WIMP-proton (WIMP-neutron) scattering  come from IceCube \cite{Aartsen:2016exj} and LUX \cite{ Akerib:2016vxi}. The IceCube limits depend on an assumption of DM $s$-wave annihilation in the sun exclusively to a single SM final state. As we will show in section \ref{DM annihilation}, our DM annihilates in the $s$-wave to both $t\bar t$ and Higgsinos. Annihilation to Higgsinos could weaken the limits somewhat if the Higgsinos are stable, but that depends in detail on the other parameters of the model (such as $\mu$, $k_u$ and $k_d$). Here we consider the simplest case where annihilation is only to $t\bar{t}$; this will provide the ``worst case scenario" where the SD bound from IceCube is strongest. In section \ref{sec:results} we will also take into account annihilation to stable Higgsinos. 

In this section, we will recast these constraints in terms of the couplings $c_h$ and $c_Z$. 
The discussion here can be viewed as an update of the nice treatment in \cite{Cheung:2012qy} with the latest experimental results (in particular LUX).  It is worth emphasizing that these bounds on $c_h$ and $c_Z$ are quite model independent. Any WIMP DM that couples to SM mainly through Higgs and $Z$ (including MSSM neutralinos) should satisfy these bounds.

\begin{figure}[t]
\centering  
\subfigure{\label{fig:feynman DD h}\includegraphics[clip, trim=2in 7.5in 3in 1in,width=0.5 \textwidth]{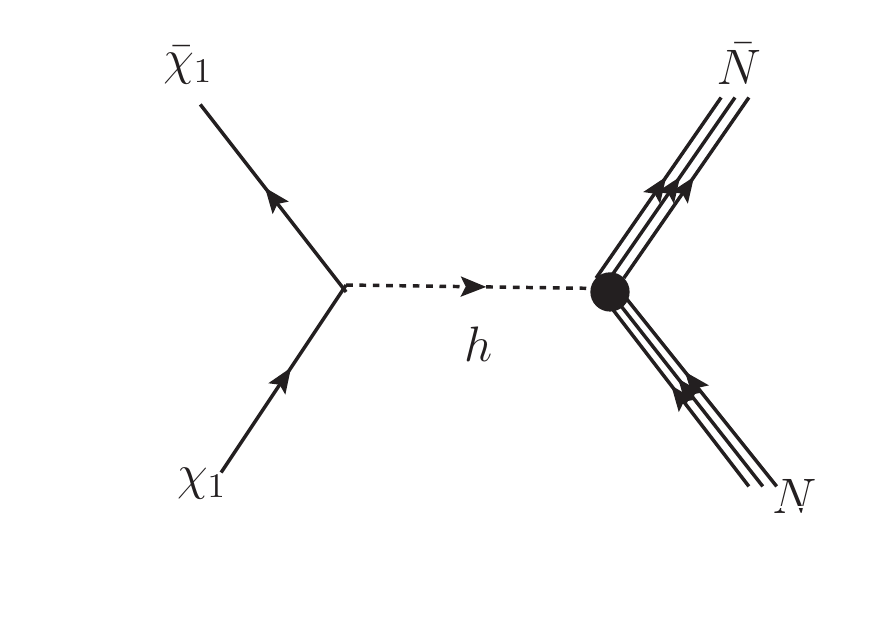}}%
\subfigure{\label{fig:feynman DD Z}\includegraphics[clip, trim=2in 7.5in 3in 1in,width=0.5 \textwidth]{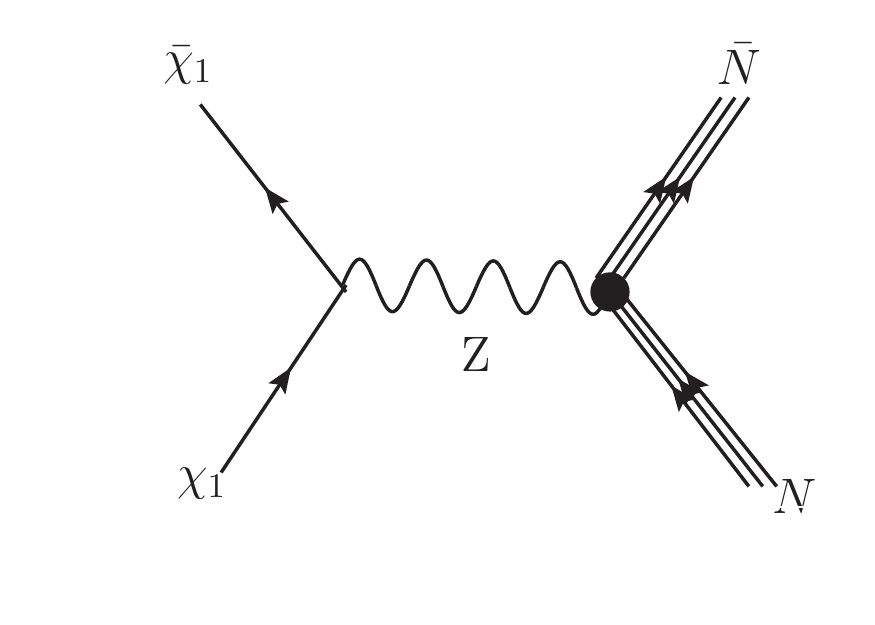}}
\caption{Higgs ($Z$) exchange diagrams contributing to SI (SD) cross sections}
{\label{fig:feynmanDD}}
\end{figure}

To convert the results of these experiments into bounds on $c_Z$ and $c_h$, we first translate $c_Z$ and $c_h$ into the couplings appearing in the  effective Lagrangian for direct detection:
\bea\label{eq:dm quark lagrangian}
\delta\mathcal{L}\supset \sum_q\Bigg(  \xi_q^{SI}(\bar\psi_\chi\psi_\chi)(\bar{q} q) + \xi_q^{SD}(\bar{\psi}_\chi \gamma^\mu \gamma_5 \psi_\chi)(\bar{q}\gamma_\mu \gamma_5 q)  \Bigg).
\eea
In Higgs and $Z$-portal DM models, the SI (SD) terms arise from Higgs  ($Z$) exchange, as shown in fig.~\ref{fig:feynmanDD}. The coefficients of the effective operators are given in terms of $c_h$ and $c_Z$ as:
\beq\label{eq:SISDcoupling}
\xi_q^{SI}= y_q \frac{c_{h} }{m^2_h} \,\, ,\qquad \xi_q^{SD} =  \frac{g_2\eta_q }{4c_W} \frac{ c_{Z} }{ m_Z^2} 
\eeq
with $y_q$ being the Yukawa coupling and $\eta_q=1$ $(-1)$ for down-type (up-type) quarks.

Then we use standard formulas that relate the DM-nucleon cross sections to $\xi_q^{SI,SD}$   (see appendix \ref{appendix:DD_to_ch/cZ} for our conventions and parameter choices).
The result, assuming $m_{DM}\gg m_{p,n}$ is given by: 
\bea\label{eq:numericalDD}
& \sigma^{SI}=c_h^2 \times(2.11\times 10^3~\text{zb})\\
& \sigma^{SD}_p=c_Z^2\times(1.17 \times10^{9}~\text{zb})\\
& \sigma^{SD}_n=c_Z^2\times(8.97 \times10^{8}~\text{zb})
\eea
In principle,  $\sigma^{SI}_p$  and $\sigma^{SI}_n$ are slightly different but the difference is negligible, so we only take  $\sigma^{SI}_p$ to represent both.

\begin{figure}
\centering
\includegraphics[scale=0.85]{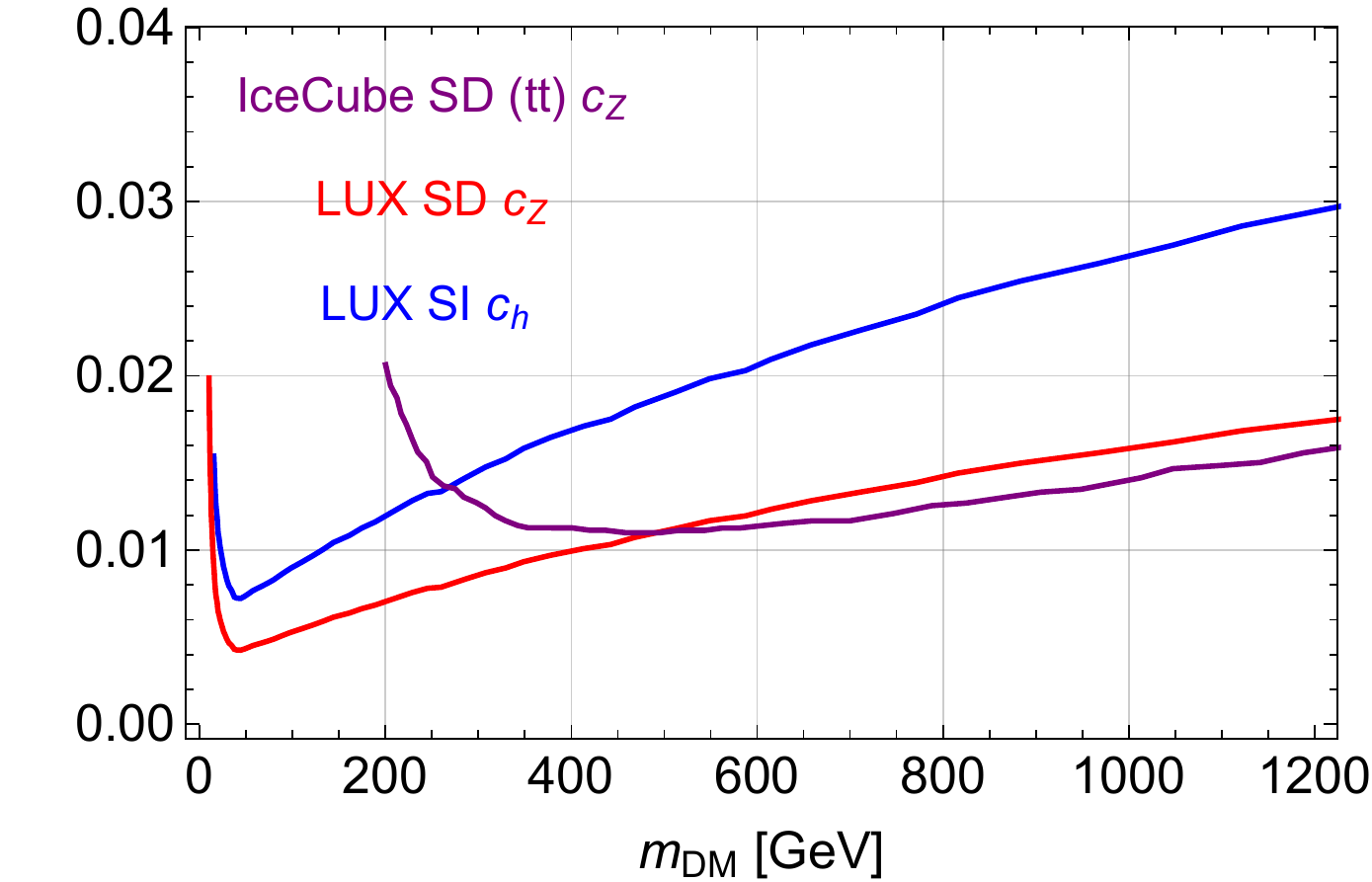}
\caption{\small{  Limits on $c_h$ from LUX \cite{ Akerib:2016vxi} (blue) and $c_Z$ from LUX \cite{ Akerib:2016vxi} (red) and IceCube $t \bar{t}$  \cite{Aartsen:2016exj} supposing DM only annihilates to $t\bar{t}$ (purple).}}
\label{fig:czch}
\end{figure}

The resulting limits on $c_h$ and $c_Z$ are shown in fig.~\ref{fig:czch}.\footnote{We agree with the limits from \cite{Cheung:2012qy} after taking into account a factor of 2 in both $c_Z$ and $c_h$ from 4-component vs.\ 2-component notation. We also agree with limits on operators from \cite{DelNobile:2013sia} modulo a factor of 4 between Dirac and Majorana fermions and a factor of a few difference between \cite{DelNobile:2013sia} and the latest LUX bounds. We do not agree with the limits on $g_A$ (related to our $c_Z$ via  $g_A = \frac{c_w}{g_2} c_Z$)  reported in fig.~3 of \cite{deSimone:2014pda}. Their limit on $g_A$, derived from essentially the same LUX results, is over an order of magnitude weaker than ours.}  
Amusingly, we note that although the constraint on the SI cross section is $\sim10^5$ stronger than the SD cross-section, translated constraints on $c_h$ and $c_Z$ are of the same order of magnitude. This is because the Higgs-nucleon effective Yukawa coupling ($y_{h N N }$) is much weaker than the $Z$-nucleon effective coupling ($\sim g_2$). Recall that the Higgs-nucleon coupling is mainly due to Higgs-gluon-gluon loop-induced interaction with heavy quarks running in the loop
\bea
y_{h N N } = \frac{\sqrt{2} \alpha_s N_H}{24 \pi v} \langle N | G^{a\mu\nu}G_{\mu\nu}^a|N\rangle=\frac{\sqrt{2} N_H m_N}{3 b v}\simeq 10^{-3}
\eea
where $N_H = 3$ is the number of heavy quarks and $b = 11-\frac{2}{3} N_H$ comes from QCD beta function at one loop. The second equality can be calculated using QCD scale anomaly that relates the QCD beta function to nucleon mass (see \cite{Gunion:1989we} for the original references).

\section{Higgs mass and Fine-Tuning}\label{Higgs mass and Fine Tuning}

In this section we will describe our calculation of the Higgs mass in the model and its implication for the fine-tuning of the EW scale. As described in the introduction, we used SARAH 4.5.8 \cite{Staub:2013tta} and SPheno 3.3.7 \cite{Porod:2003um,Porod:2011nf}  to include all the loop corrections (contributions up to two loops both from the MSSM and the dark sector \cite{Goodsell:2014bna,Goodsell:2015ira}). Here we will describe an analytic treatment of the dominant one-loop contributions from $k_u$ and $k_d$. This will serve as a valuable source of intuition, as well as a validation of the full two-loop numerical calculation (for more details on the validation, see appendix \ref{appendix:Validating_Higgs_mass}).

The one-loop Higgs mass  was previously computed in the literature using the Coleman-Weinberg potential in closely-related vector-like extensions of the MSSM \cite{ Babu:2004xg, Babu:2008ge, Martin:2009bg,  Martin:2010dc}. However, there are some key differences with our case that necessitate a fresh look. First, as noted above, in these past works, the vector-like extension was Dirac, while ours is Majorana (the difference between $W\supset M S\tilde S$ and $W\supset {1\over2}MS^2$). This leads to small differences in the formula for the Higgs mass. Second, previous works presented analytic formulas for the one-loop Higgs mass only in the simplified limit with common fermion masses ($M_L=M_S$). Motivated by the DM side of the story, we will need the Higgs mass in a rather different regime, the mostly-singlet regime where $M_S\ll M_L$. 

Other effects that we will ignore in our discussion here, but that are taken into account in the full numerical SARAH-SPheno calculation, include $g_{1,2}^2$ corrections, two-loop corrections, and the effective $A$-terms due to $\mu$. The effects of $g^2$ are about a 10-20\% correction to $\delta m_h^2$, which amounts to a 2 GeV shift in $m_h$. That matters for our calculations quantitively but not qualitatively. The $\mu$ values we consider in this paper motivated by naturalness are small enough that $\mu$ has a negligible effect on the Higgs mass. Finally, we are interested in moderately-large values of $\tan\beta$ (e.g.\ $\tan\beta=10$) but for simplicity we will present the $\tan\beta\to\infty$ limit here. The corrections due to $1/\tan\beta$ also do not make a qualitative difference. (In particular, there are no blind-spot cancellations here.) 

\begin{figure}
\centering 
\includegraphics[width=\textwidth]{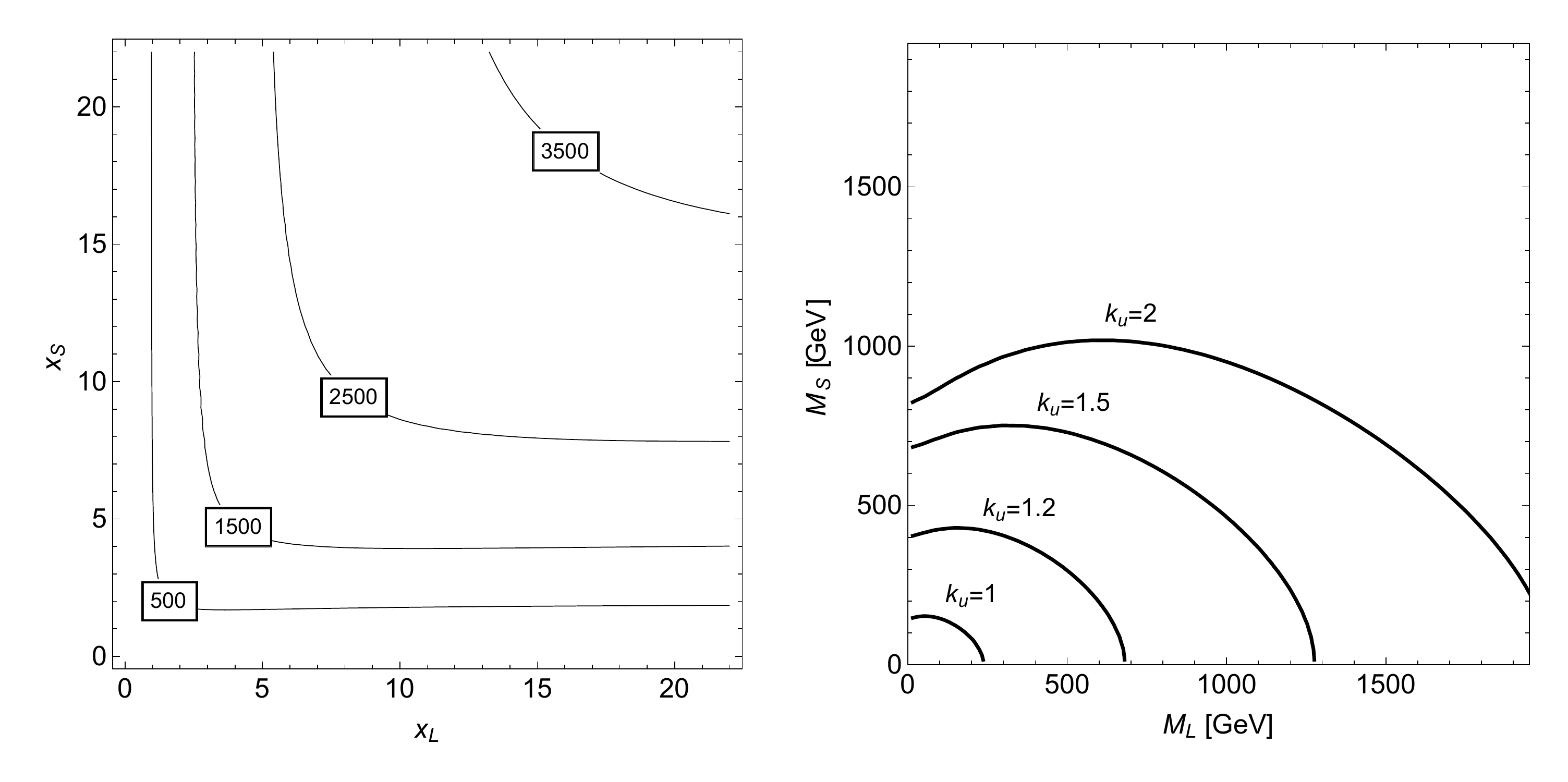}
\caption{\small{ Left: contours of $k_u^{-4}\delta m_h^2$ in GeV$^{2}$ in the $(x_L, x_S)$ plane, according to (\ref{eq:deltamhsq}). Right: contours of $\Delta=20$ in the $(M_L, M_S)$ plane for different values of $k_u$ that have a Higgs mass at 125 GeV.}}
\label{fig:mhsq}
\end{figure}

With all of these simplifying assumptions, the result of our one-loop Coleman-Weinberg calculation is:
\beq\label{eq:deltamhsq}
\delta m_h^2 = {1\over 4\pi^2} k_u^4 v^2\left( f_1 \log (1+x_L^2)
+ f_2\log (1+x_S^2) + f_3 \log {x_S^2\over x_L^2}\right)
\eeq
with
\bea
& f_1 = { (2 x_L^4 + x_L^6 + 3 x_L^2 x_S^2 + 3 x_L^4 x_S^2 - x_S^4)x_S^2\over (x_L^2-x_S^2)^3}\\
& f_2 = {(x_L^2-5x_S^2-x_L^2x_S^2-3x_S^4)x_L^4\over (x_L^2-x_S^2)^3}\\
& f_3 = {x_L^4x_S^2(x_L^2+3x_S^2)\over (x_L^2-x_S^2)^3}
\eea
where $x_L=m/M_L$ and $x_S=m/M_S$. 
  A plot of $k_u^{-4}\delta m_h^2$ is shown in fig.~\ref{fig:mhsq} (left). We see that $\delta m_h^2$ asymptotes to a finite value as $x_L\to\infty$ or $x_S\to \infty$. In these limits (corresponding to mostly-doublet and mostly-singlet DM respectively), the dependence  on the DM mass drops out, and $\delta m_h^2$ is controlled by the ratio of the soft mass to the  heavier mediator scale ($M_S$ or $M_L$ respectively).
  
To raise the Higgs to 125~GeV in this paper, we rely on a combination of the extra vector-like matter and MSSM stops. For stops at 1~TeV, which satisfy the current experimental bounds and imply about a $\sim 10$\%  tuning of the EW VEV, the MSSM contribution to the Higgs mass is about 110~GeV (for a recent review see e.g.\ \cite{Draper:2016pys}). Therefore the target for $\delta m_h^2$ from the dark sector is:
\beq\label{eq:deltamhsqtarget}
\delta m_h^2 \approx 3500~{\rm GeV}^2
\eeq
This selects out a contour in the $(x_S,x_L)$ plane as shown in fig.~\ref{fig:mhsq} (left), according to the value of $k_u$.

This has the following implications for the fine-tuning of the EW scale. Just as the dark sector lifts the physical Higgs mass analogous to stops in the MSSM, it also contributes to the fine-tuning of the EW scale through the renormalization of $m_{H_u}^2$. Following \cite{Hall:2011aa,Kitano:2005wc}, we define the measure
of fine-tuning to be:
\bea\label{eq:FT1}
\Delta &= \frac{2 \delta m_{H_u}^2}{m_h^2}
\eea
where $\delta m_{H_u}^2$ is the running of $m_{H_u}^2$ due to the new fields
\bea\label{eq:FT2}
\delta m_{H_u}^2 &= \frac{k_u^2 m^2}{8 \pi^2} \log{\frac{\Lambda_{UV}}{\Lambda_{IR}}}  
\eea
Optimistically we take $\Lambda_{UV} = 10\, \Lambda_{IR} \sim 10\, \text{TeV}$. We can combine this with (\ref{eq:deltamhsq}) and (\ref{eq:deltamhsqtarget}) as follows. For a given value of $k_u$ and a given point in the $(M_S, M_L)$ plane, we can solve (\ref{eq:deltamhsqtarget}) for the soft mass $m$. Then substituting this into (\ref{eq:FT1}), we get a value for $\Delta$. Regions of $\Delta\le 20$ are shown in  fig.~\ref{fig:mhsq} (right) for different representative values of $k_u$. We see that we need $k_u\gtrsim 1$ to have any viable parameter space at all for a natural SUSY 125 GeV Higgs. This is not surprising, since from (\ref{eq:deltamhsq}), we see that $k_u$ plays the role that $y_t$ plays for the MSSM stops. Of course, corrections we have neglected such as the $D$-terms and two-loop effects will modify this quantitatively. However, we will see that the same qualitative implications for fine-tuning and $k_u$ will persist in our final plots.


\section{The need for mostly-singlet DM}{\label{sec:mostly_singlet}}

In section  \ref{sec:model}, we derived formulas for $c_h$ and $c_Z$ in terms of the parameters of the model, while in section \ref{sec:DMDD} we showed that direct detection limits on $c_h$ and $c_Z$ are at the $\CO(10^{-2})$ level. Finally, in section \ref{Higgs mass and Fine Tuning}, we argued that we need $k_u\gtrsim 1$ in order to have any viable parameter space for a natural SUSY Higgs at 125 GeV. Here we will combine these facts and show that the DM must be mostly singlet in order to be consistent with all the constraints.

Basically there are three possibilities: the well-tempered regime where $|M_L-M_S| \lesssim v$ (recall our convention is that $M_S$ and $M_L$ are positive), the mostly-doublet regime where $M_L<M_S$ and $v\ll M_S$, $M_S-M_L$, and the mostly-singlet regime where $M_S<M_L$ and $v\ll M_L$, $M_L-M_S$. Keeping in mind that we need $k_u\gtrsim 1$ and large $\tan\beta$ for a natural Higgs mass, the challenge is to decrease $c_h$ and $c_Z$ to the $10^{-2}$ level. In fact, $c_h$ alone is enough to rule out all but the mostly-singlet case. We will comment on the implications for $c_Z$ in sections \ref{DM annihilation} and \ref{sec:results}.

Examining the formula for $c_h$ (\ref{eq:chdef}), we see that for $|M_L-M_S|\sim v$ and $\hat k_{u}\sim 1$, we have $c_h\sim\CO(1)$. (In particular, there is a cancellation in the denominator, leaving it $\CO(v M_S)$). This rules out the well-tempered case. 

The mostly-doublet case is ruled out separately by two independent considerations. First, from fig.~\ref{fig:mhsq}, we see that in order to be natural and mostly-doublet, we must have the DM mass below $\sim 800$~GeV. However, we know by analogy with pure Higgsinos in the MSSM that the thermal relic density constraint requires $M_L\ge 1$~TeV. (The mostly-doublet DM in this model has additional annihilation modes due to $k_u$ and $k_d$, so $M_L$ will be even larger.) So the mostly-doublet scenario is not promising for naturalness.

Also, from direct detection, we are basically forced into the mostly-singlet regime. In order to lower $c_h$ by two orders of magnitude, we must either (a) raise $M_L$ or $M_S$ to increase the denominator of (\ref{eq:chdef}), or (b) cancel the two terms in the numerator of (\ref{eq:chdef}). 
\begin{enumerate}[label=(\alph*)]

\item Increasing the denominator of  (\ref{eq:chdef}) necessitates either $M_L$ or $M_S\gg v$. In the former, corresponding to mostly-singlet DM, we see that $c_h\propto 1/M_L^2$ and we can achieve the required level of suppression for $M_L\sim 1-2$~TeV for $M_S\sim v$ and $k_u\sim 1$. Meanwhile for the latter, corresponding to mostly-doublet DM, we see that $c_h\propto 1/M_S$ and therefore much larger $M_S\sim 2-5$~TeV is required for $M_L\sim v$ and $k_u\sim 1$. The latter is greatly disfavored by naturalness (it would likely be as fine-tuned as 10~TeV stops in the MSSM).

\item Cancelling the two terms in the numerator requires 
\beq
{M_L\over m_\chi}\sim -{1\over2}{k_u\over k_d}\tan\beta
\eeq
This is the blind spot. Since $k_u\gtrsim 1$ and we are in the large $\tan\beta$ limit, the RHS is generally much greater than one for any reasonable value of $k_d$. Therefore we must be in the mostly-singlet DM regime to realize the blind spot.

\end{enumerate}

\noindent We conclude that several different constraints independently point at mostly-singlet DM as the only viable possibility.

\begin{figure}
\centering
\includegraphics[scale=0.85]{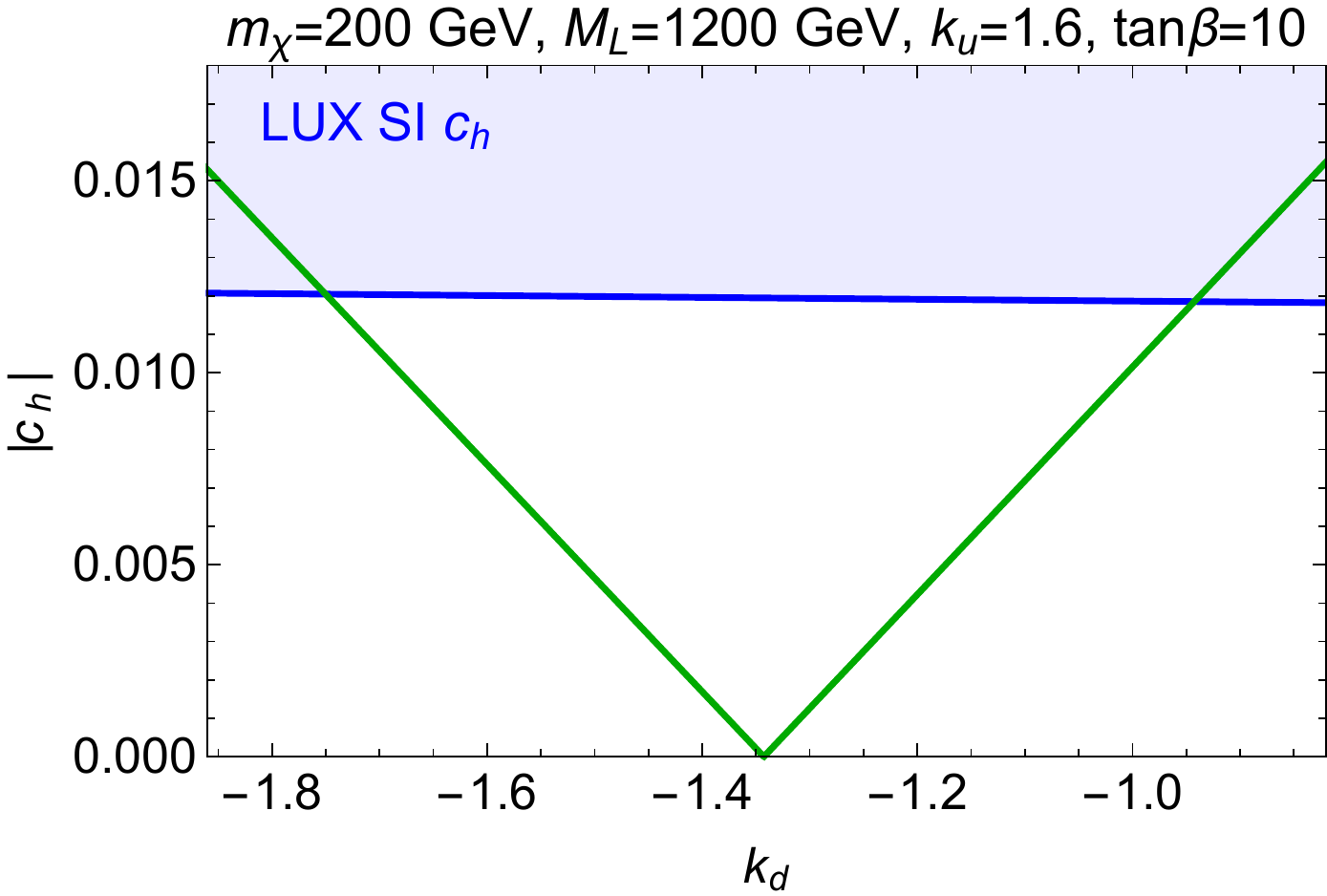}
\caption{\small{Values of the coupling $c_{h}$ while varying $k_d$ for a sample point of the parameter space. The values on the shaded area are excluded by LUX. }}
\label{fig:plotBS}
\end{figure}

For later reference we exhibit $c_h$ and $c_Z$ in the mostly-singlet limit 
\bea\label{eq:chcz}
c_h &= -{m_\chi+ {2k_d M_L\over k_u \tan\beta}\over\sqrt{2}v} {k_u^2v^2\over M_L^2} + \dots\\
c_Z &= -{g_2\over4c_W}{k_u^2v^2\over M_L^2}+\dots
\eea
Here we have taken $M_L\to \infty$ and $\tan\beta\to\infty$ holding fixed $M_L/\tan\beta$ and all the other mass scales. In fig.~\ref{fig:plotBS}, we exhibit the amount of blind spot cancellation that is required by the SI bounds, for a typical choice of parameters that will lead to a viable relic density. We show this behavior by varying $k_d$ keeping other parameters fixed. We can see that we need only a very mild cancellation to satisfy the constraint on $c_{h}$. Most of the suppression of $c_h$ is coming from large $M_L$, which as we will see in the next section is fixed by the thermal relic abundance constraint.

In the same mostly-singlet regime, we also exhibit $\delta m_h^2$: 
\beq \label{eq:deltamhsqsimp}
\delta m_h^2 = {k_u^4v^2\over 4\pi^2} \log(1+x_L^2) - {3  k_u^4v^2 x_L^4\over 4\pi^2 x_S^2}\log {1+x_L^2\over x_L^2} + \CO(x_S^{-4})
\eeq
As noted in the previous subsection, the Higgs mass in this limit to leading order does not depend at all on the DM mass $M_S$. So the Higgs mass constraint to leading order in the mostly-singlet regime becomes a constraint on $k_u$ and $m/M_L$. For example, according to (\ref{eq:deltamhsqsimp}), in order to achieve $\delta m_h^2=3500$~GeV$^2$ for $k_u =1.6$, we need $x_L \approx 1$.


\section{DM annihilation in the mostly-singlet regime}\label{DM annihilation}

An attractive feature of WIMP dark matter is its potential to naturally explain the observed relic abundance via the thermal freeze-out mechanism. Following the usual procedure (see e.g.\ the classic review  \cite{Jungman:1995df}), we have
\bea\label{eq:omegadm}
\Omega_{DM} h^2 \approx 9.2\times10^{-12}\,\,{\rm GeV}^{-2} \times \left(\int_{x_f}^\infty dx\, {\langle \sigma v_\chi\rangle\over x^2}\right)^{-1}
\eea
The integral over $x$ takes into account annihilation after freeze-out, and $x_f=m_\chi/T_f\approx 25$ parametrizes the freeze-out temperature. $\langle \sigma v_\chi\rangle$ is the thermally-averaged DM annihilation cross section $\chi\chi\to XY$, summed over all final states $X$ and $Y$. This  is usually expanded in the small velocity limit:
\bea
\sigma_{XY} \,v_\chi = r_{XY}(a_{XY} + b_{XY} \,v_\chi^2 + \O(v_\chi^4)),
\eea
where $r_{XY}\equiv \sqrt{1-(m_X+m_Y)^2/4m_\chi^2}$ is a kinematic phase space factor. At the time of freeze-out, the DM relative velocity is typically $v_\chi^2\sim 0.1$. Therefore, the annihilation cross section is generally controlled by the $s$-wave contributions $a_{XY}$, unless they are suppressed for some reason. 

In our model,  the dark matter has many interactions and annihilation channels that should all be considered in full generality. As described in the introduction, for numerical calculations we use micrOMEGAs 4.1.8 \cite{Belanger:2014vza}  source code generated by SARAH 4.5.8 \cite{Staub:2013tta} to accurately take these into account. However in the mostly singlet limit that we are interested in, the cross sections simplify and we can have an analytic understanding of the behaviour of our model. We will assume that DM is lighter than all MSSM superpartners except possibly the Higgsinos, which are forced to be light $\mu\sim v$ by naturalness. In this case, the freeze-out process happens only through annihilation to SM particles and the Higgsinos. Including the Higgsinos in the story is a major difference from simplified-model-analyses of singlet-doublet dark matter, which generally just add the singlet and doublets to the SM. As we will see, the Higgsinos can be a major part of the DM annihilation in the early universe.

The full cross sections are too complicated to print here. Instead, we will expand in the mostly-singlet limit $M_S<M_L$, $v\ll M_L$, $M_L-M_S$ with the further assumption that $v\ll M_S$. This suffices for our purposes and results in relatively simple expressions. (One exception is the tree-level, $s$-wave $t\bar t$ cross section in the next subsection, for which we can write down an extremely simple exact expression in terms of $c_Z$.)

\subsection{DM annihilation to fermions}

The fermions have $s$-wave contributions
\bea\label{eq:difermions}
& a_{f\bar f}=  {3 k_u^4 \over 32\pi} { m_f^2\over  M_L^4 (1-\epsilon^2)^2 }\\
& a_{\psi_H\psi_H}= {(k_d^2+k_u^2)^2\over 16\pi }
{\mu^2  \over M_L^4(1+\epsilon^2+x_L^2)^2} \\
\eea
where $\epsilon\equiv M_S/M_L$, and $x_L\equiv m/M_L$ was defined in section \ref{Higgs mass and Fine Tuning}. In the second line, we have summed over the various Higgsino final states including both neutralinos and charginos, assuming a pure MSSM Higgsino (i.e.\ $M_{1,2}$ decoupled). The fermion  $b$ coefficients are always subdominant (suppressed by both $v_\chi^2$ and $v^2/M_L^2$), so we have not included them here.

The fermion cross sections are all suppressed by the square of the fermion mass, so $t\bar t$ and Higgsinos are the dominant channels. This is the famous $s$-wave helicity suppression of DM annihilation to fermion pairs. 

Although $t\bar t$ and Higgsinos are parametrically similar, their diagrammatic origin is entirely different. The former (latter) arise from $s$-channel $Z$ ($t$-channel superpartner) exchange. As a result, the Higgsinos are suppressed by the soft mass $m$. For $k_u=1.6$, we saw in section \ref{sec:mostly_singlet} that we need $x_L\approx 1$ for $m_h=125$~GeV, so the suppression is not large. Also, $\mu$ is constrained to be $\lesssim 300$~GeV by naturalness. So all in all, the Higgsino contribution ends up generally of the same order or smaller than $t\bar t$. 

The fact that the SM fermions all arise from $s$-channel $Z$ diagrams means that they have a simple exact expression {\it beyond} the small $v$ approximation:
\beq\label{eq:affbarcZ}
 a_{f\bar f}=c_Z^2{3y_f^2 \over 4\pi m_Z^2 } \\
\eeq
In other words, $c_Z$ controls both the SD direct detection cross section and the annihilation to $t\bar t$. Therefore, we expect to see a fairly direct correlation between the SD direct detection limits and the relic density constraint.

\subsection{DM annihilation to bosons}

Meanwhile the diboson cross sections are all $p$-wave to leading order:
\bea\label{eq:dibosons}
 & b_{hh} = b_{ZZ}= {  k_u^4   \over 384\pi }{\epsilon^2(3+2\epsilon^2+3\epsilon^4)
\over M_L^2(1+\epsilon^2)^4} \\
 & b_{hZ}={  k_u^4  \over 96\pi }{\epsilon^2\over M_L^2 (1+\epsilon^2)^2} \\
 & b_{WW} =2 b_{hh}+b_{hZ} \\
\eea
Here we took $\tan\beta\to\infty$ for simplicity; we checked that the $1/\tan\beta$ corrections are irrelevant. The $s$-wave contributions are suppressed by $v^4/M_L^4$ so they are always subdominant to the $p$-wave contributions shown here.

Clearly, the diboson cross sections exhibit some interesting features. They are nonvanishing even in the $v\to 0$ limit, so they can be understood as a consequence of $SU(2)_L\times U(1)_Y$ symmetry. These tree-level cross sections arise entirely due to the longitudinal components of the $W^\pm$ and $Z$ bosons, which by the Goldstone equivalence theorem are also equivalent to the charged and neutral Goldstones $G^\pm$ and $G^0$ respectively. Under a $U(1)_Y$ rotation, $h\to G^0$ and $G^0\to -h$, while under an $SU(2)_L$ rotation, $W^\pm\to h\pm i G^0$. This explains both relations in (\ref{eq:dibosons}).

Comparing $t\bar t$ and Higgsinos to the total diboson cross section, we see that parametrically the latter can be larger than the former, for sufficiently large $M_L$. However the cross over point is generally at very large $M_L$ and $M_S$. For instance, for $\epsilon=1/2$ and $x_L=1$, we find the cross over to be in the range $M_L\sim 2.7-3.6$~TeV for $\mu\sim 100-300$~GeV.  This is well beyond the naturalness-motivated part of the parameter space that we are focusing on in this paper. Therefore we conclude that the total $\sigma v_{\chi}$ is always dominated by $t\bar t$ and Higgsinos, and dibosons are always a subdominant part of it.

\subsection{Total annihilation cross section}
\label{sec:totalsigmav}

\begin{figure}[t]
\centering
\includegraphics[scale=0.6]{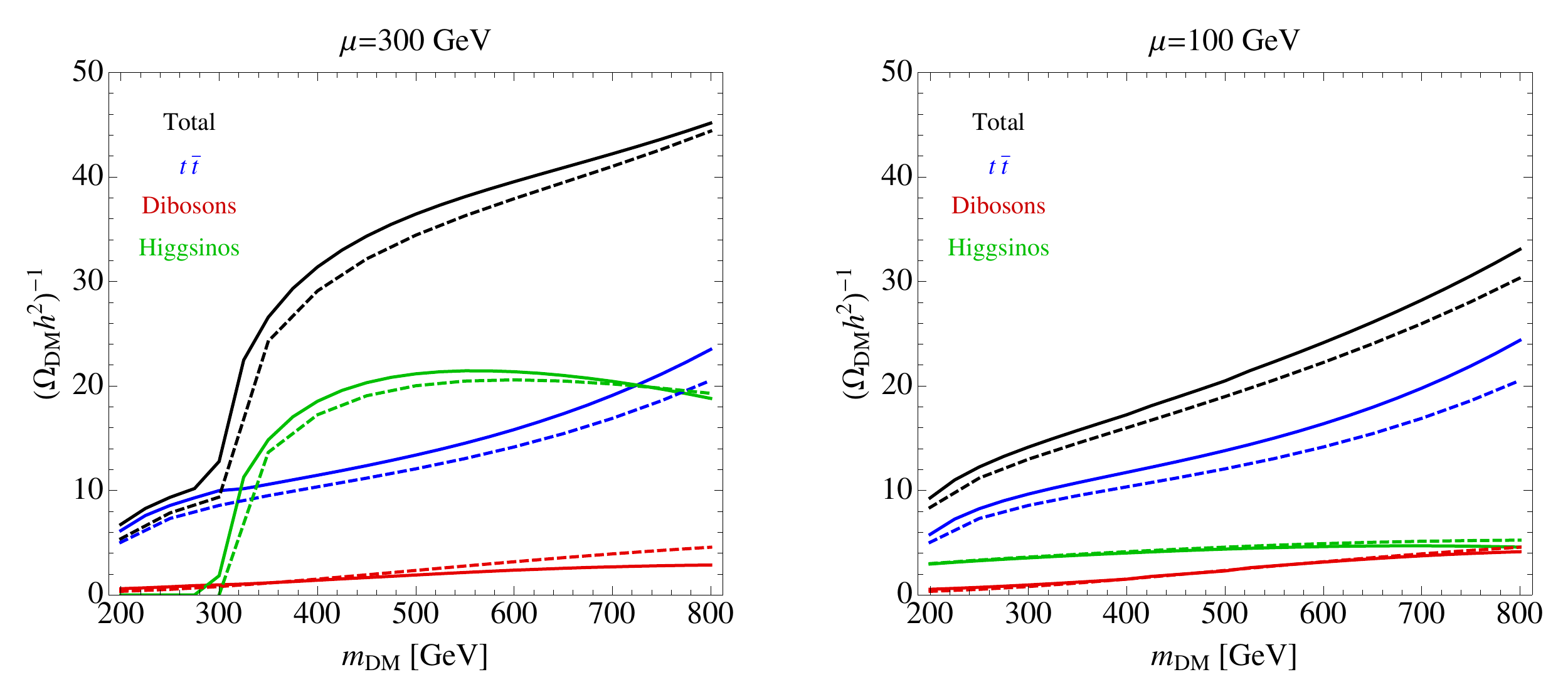}
\caption{\small{Inverses of the total relic abundance (black) as well as the individual contributions from $t\bar t$ (blue) and dibosons (red) as calculated numerically by micrOMEGAs 4.1.8 (solid) and the analytic equations (\ref{eq:difermions}) and (\ref{eq:dibosons}) (dashed), for $M_L=1.2$~TeV, $k_u=1.6$, $k_d=-1.5$ and $\tan\beta=10$.} }
\label{plot:MOvsanalytic}
\end{figure}

We have shown analytically  that the relic density is dominated by $s$-wave annihilation to $t\bar t$ and Higgsinos (assuming of course that the DM is above the respective thresholds):
\beq\label{eq:sigmatotapprox}
\sigma v_\chi \approx a_{t\bar t}+a_{\psi_H\psi_H} =  {3 k_u^4 \over 32\pi} { m_f^2\over  M_L^4 (1-\epsilon^2)^2 }+ {(k_d^2+k_u^2)^2\over 16\pi }
{\mu^2  \over M_L^4(1+\epsilon^2+x_L^2)^2}
\eeq
A plot comparing our analytics to micrOMEGAs is shown in fig.~\ref{plot:MOvsanalytic} for fixed choices of the parameters; we see there is excellent agreement across the entire range of relevant DM masses. We  confirm that the dibosons are never more than $\sim 10$\% of the relic density across the entire parameter range of interest. Higgsinos and $t\bar t$ are comparable for $\mu\sim 300$~GeV, while for $\mu\sim 100$~GeV, $t\bar t$ dominates, as expected from the $\mu$ dependence of the Higgsino cross section (\ref{eq:difermions}).

One very interesting consequence of (\ref{eq:sigmatotapprox}) is that in the limit of large $M_L$, the DM mass drops out of the annihilation cross section.  Furthermore, we have seen that we need $k_d\sim k_u$ for the blind spot,  $x_L\approx 1$ for the Higgs mass, and $\mu\sim m_t$ for naturalness. Thus the WIMP miracle transforms from being a constraint on the WIMP mass to being a constraint on the mediator scale $M_L$! This helps to relieve the ``WIMP little hierarchy problem", whereby the preference of the thermal relic constraint for TeV-scale WIMPs is in tension with naturalness. Comparing with (\ref{eq:chcz}), we also expect that the relic density constraint will essentially fix $c_Z$ to a unique value. We will confirm this in the next section with our full numerical scans and discuss its implications for SD direct detection.

\section{Putting it all together}\label{sec:results}

\subsection{Plots in the $M_L$-$M_S$ plane}

Having described the various individual components of the  analysis of the model (direct detection, the Higgs mass, and the relic abundance), we will now combine them and describe how the different constraints interact to produce the viable parameter space of the model.

\begin{figure}[t]
\centering
\includegraphics[width=0.95\textwidth]{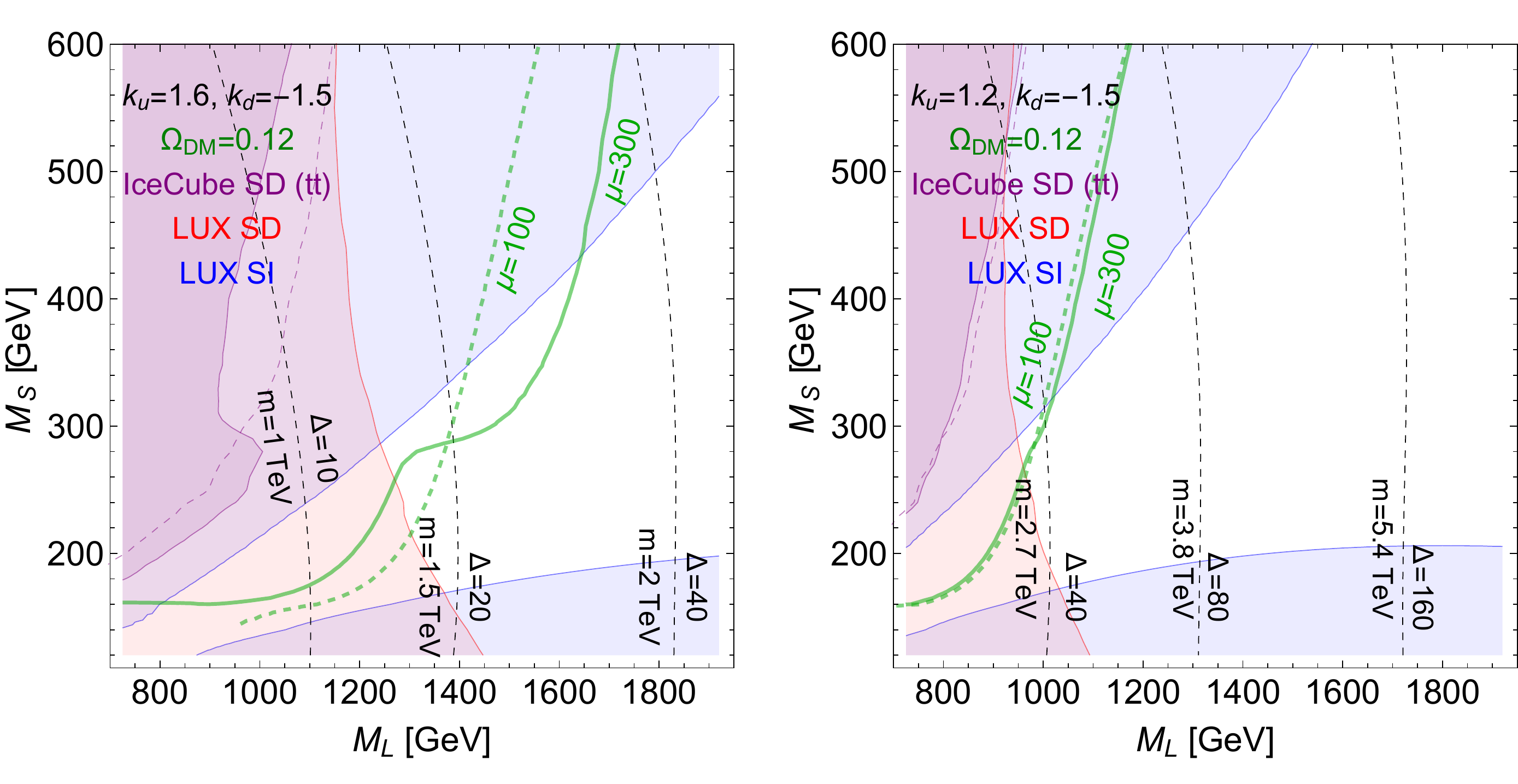}
\caption{\small{$M_L-M_S$ parameter space scan with micrOMEGAs 4.1.8 \cite{Belanger:2014vza} for $k_u=1.6$, $k_d=-1.5$ (left) and $k_u=1.2$, $k_d=-1.5$ (right), $\tan \beta=10$ and all MSSM soft masses  and gauginos at 1 TeV. We show the exclusion regions from $\s^{SI}$ in blue and $\s^{SD}_n$ in red, both from LUX \cite{Akerib:2016vxi} and for $\s^{SD}_p$ from IceCube $t\bar{t}$ with $\mu=300$ GeV ($\mu=100$ GeV)\cite{Aartsen:2016exj} in solid (dashed) purple. The dashed black lines are the fine-tuning contours for $\mu=300 \,\text{GeV}$.
The contour of $\Omega_{DM}=0.12$ with $\mu=300$ GeV ($\mu=100$ GeV) is in solid (dashed) green.}}
\label{fig:plotmoneypaper}
\end{figure}

In fig.~\ref{fig:plotmoneypaper} we show contour plots for numerical scans over the ($M_L,M_S$) plane for fixed values of $k_u,\,k_d$ and $\mu$. We choose four sets of benchmark parameters: large coupling ($k_u=1.6$, $k_d=-1.5$) and small coupling ($k_u=1.2$, $k_d=-1.5$); and large $\mu$ ($\mu=300$~GeV) and small $\mu$ ($\mu=100$~GeV).

We see the impact of the direct detection limits on the parameter space of the model. 
The LUX SI and SD limits are strongest almost everywhere except a tiny sliver for large $M_S$ in the $k_u=1.2$, $k_d=-1.5$ case where IceCube has an impact. (Note that the LUX limits assume the singlet-doublet sector comprises all of the DM, regardless of whether it is thermal or not.) The SD (SI) limits primarily cover the lighter (heavier) DM mass region. The heavier DM region is ruled out because we are holding fixed $k_d$, so as one increases $m_\chi$ the blind spot cancellation shown in (\ref{eq:chcz}) becomes less effective.

For every point in the plane we numerically solved (using SPheno) the $m_h=125$~GeV constraint for the common soft mass $m$; these contours are shown in fig.~\ref{fig:plotmoneypaper} along with their corresponding tuning. These contours are mostly vertical; as discussed in section \ref{Higgs mass and Fine Tuning}, the soft mass and $\Delta$ depend primarily on $M_L$ since the dependence on $M_S$ drops out to leading order at large $M_L$.

Finally, we used micrOMEGAs to numerically solve the thermal relic density constraint $\Omega_{DM}=0.12$  \cite{Ade:2015xua}; this fixes $M_L$ as a function of $M_S$ and these contours are shown in green for various choices of the parameters. Note the rapid increase in $M_L$ across the top and Higgsino thresholds. Here new $s$-wave annihilation channels open up, and so larger values of $M_L$ are needed to maintain the overall annihilation rate at the thermal relic value. This effect is more pronounced for larger values of $k_{u,d}$ and for larger values of $\mu$. Indeed, in section \ref{DM annihilation} we saw that the annihilation cross sections to $t\bar{t}$ and Higgsinos are enhanced for greater $k_{u,d}$, and the Higgsino cross section in particular is proportional to $(k_d^2+k_u^2)^2\mu^2$. 

Since larger $M_L$ decreases direct detection cross sections, increasing $k_{u,d}$ and $\mu$ also increases the viable parameter space for thermal relic DM. The Higgsino channels in particular allow the model to survive the direct detection limits over a wider range of parameter space than would have been the case for non-supersymmetric singlet-doublet DM.  Fig.~\ref{fig:plotmoneypaper} also shows that larger $k_{u,d}$ is better for fine-tuning, confirming our discussion in section \ref{Higgs mass and Fine Tuning}. The only potential drawback of the larger coupling choice is (as we will discuss in section \ref{sec:Landau Poles}) that the former has a lower Landau pole ($\Lambda\sim 10^2$~TeV vs $\Lambda\sim 10^3$~TeV).

\begin{figure}[t]
\centering
\includegraphics[scale=0.7]{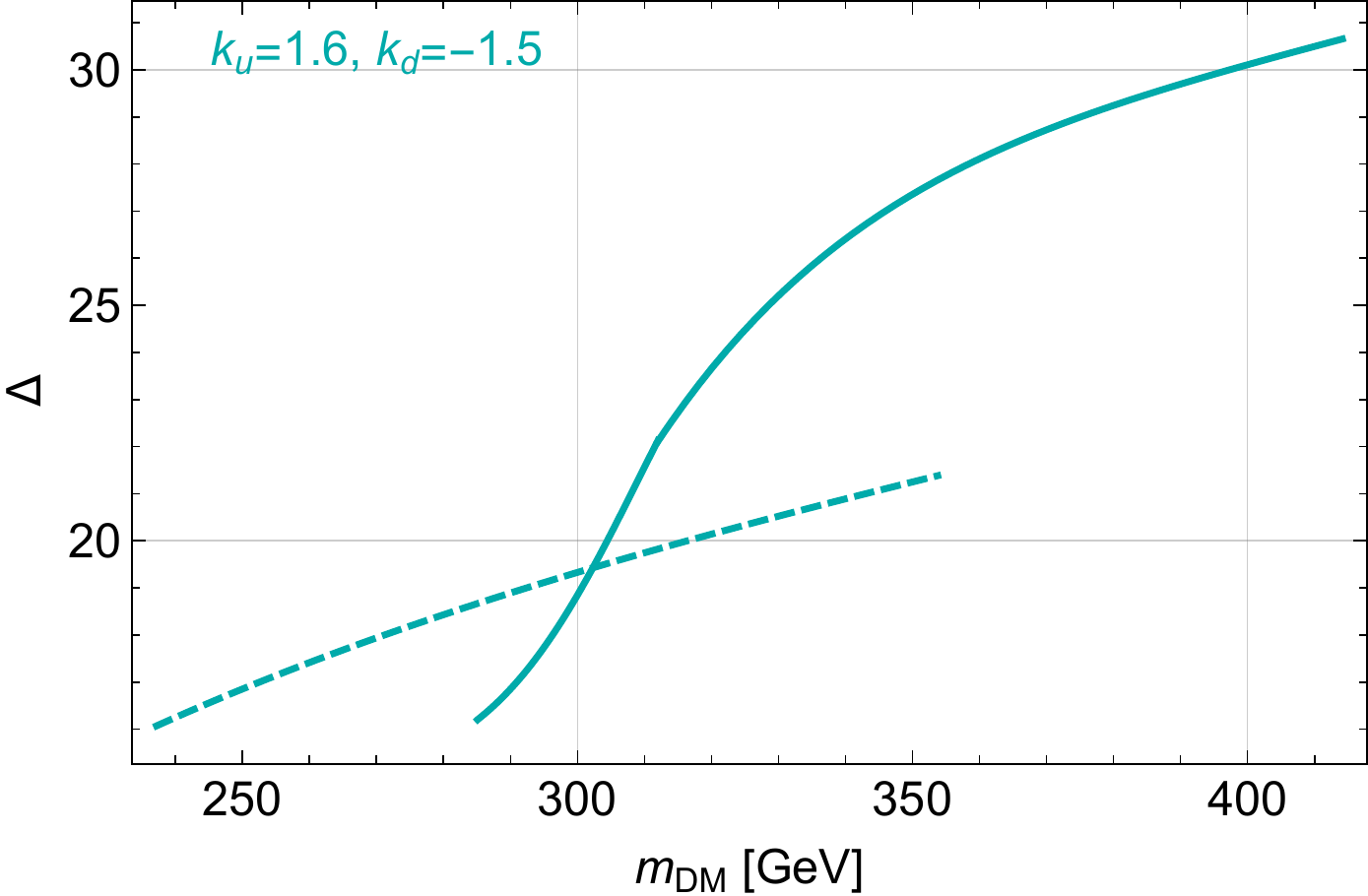}
\caption{\small{Fine-tuning for the right relic abundance contours of fig. \ref{fig:plotmoneypaper} (left) that are allowed by direct detection. We show the case for $\mu=300$ GeV (solid) and $\mu=100$ GeV (dashed).}}
\label{plot:FTOmega}
\end{figure}

 Away from the top and Higgsino thresholds, we see that the relic density contours are mostly vertical, meaning that the relic density constraint becomes a constraint primarily on $M_L$, once the other parameters ($k_u$, $k_d$, $m$, $\mu$) are fixed, i.e.\ the WIMP DM mass drops out to leading order. This confirms our analytics in the previous section.

\subsection{Projecting onto the thermal relic contour}

Finally let us impose the relic density constraint $\Omega_{DM}=0.12$ and see how various parameters vary along the green contours in fig.~\ref{fig:plotmoneypaper}. In 
fig.~\ref{plot:FTOmega} we show the fine-tuning for the points with the  correct relic abundance. It is remarkable that there are allowed regions of the parameter space with $ \Delta\lesssim20$, making this model much less tuned than the MSSM.

\begin{figure}[t]
\centering
\includegraphics[scale=0.62]{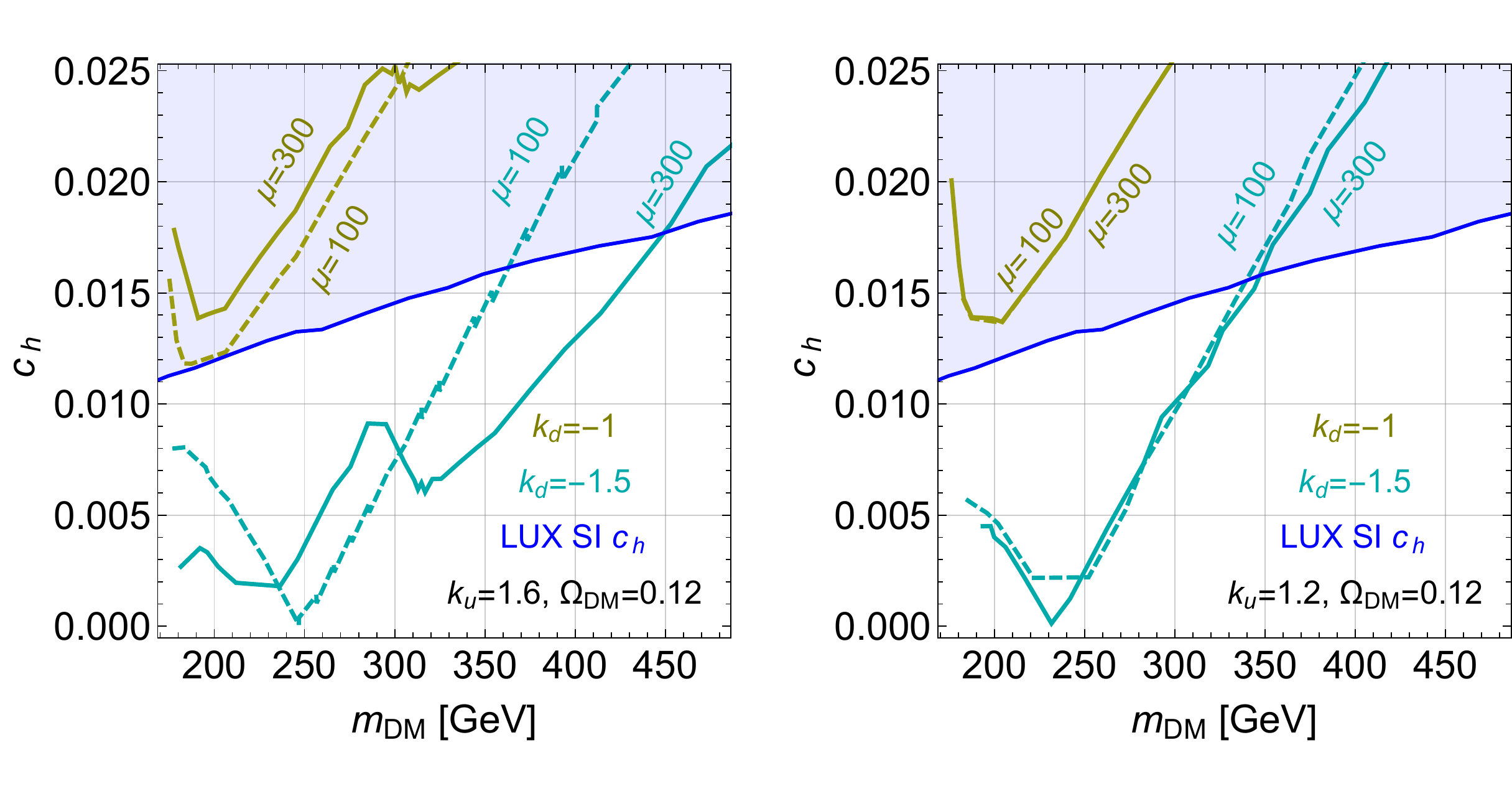}
\caption{\small{Values of the coupling $c_h$ for the points with $\Omega_{DM}=0.12$ for different values of $k_d$. We show the values for $k_u=1.6$ (left) and $k_u=1.2$   (right). The exclusion region from $\s^{SI}$ is in blue.}}
\label{plot:BSNum}
\end{figure}

In fig.~\ref{plot:BSNum} we show $c_h$ for the points of the parameter space that satisfy $\Omega_{DM}=0.12$ (including both allowed and excluded points from direct detection limits). We see that varying $k_d$ we can move toward the blind spot and satisfy the SI direct detection bounds.

\begin{figure}[t]
\centering
\includegraphics[scale=0.62]{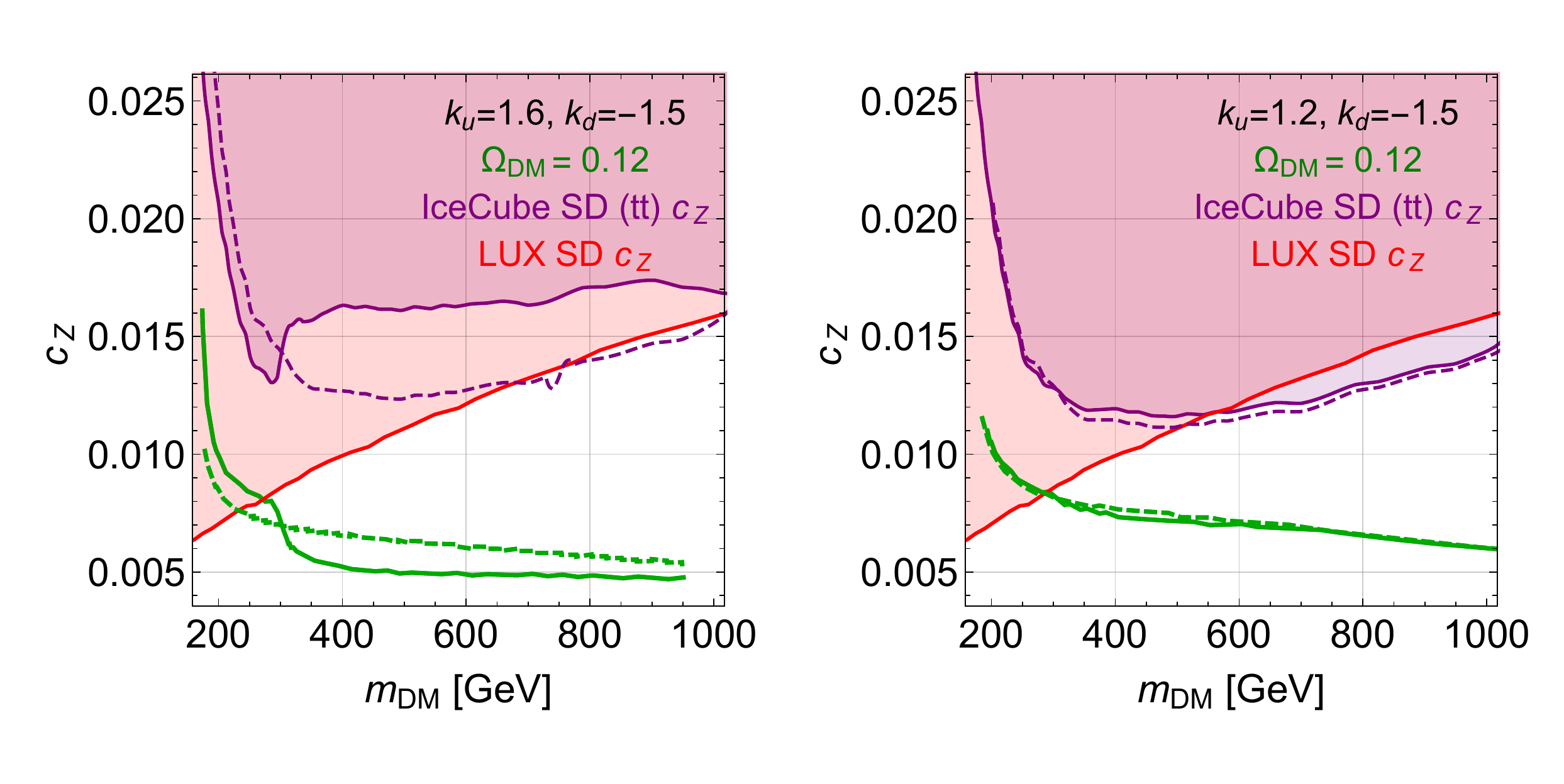}
\caption{\small{Values of the coupling $c_Z$ for the points with $\Omega_{DM}=0.12$ for $\mu = 300 \, \text{GeV}$ ($\mu = 100 \, \text{GeV}$) in solid (dashed) green. The exclusion region from $\s^{SD}_n$ is in red and from $\s^{SD}_p$ with $\mu=300$ GeV ($\mu = 100$ GeV) is in solid (dashed) purple.}}
\label{plot:plotSDOmega}
\end{figure}

Similarly, in fig.~\ref{plot:plotSDOmega} we show $c_Z$ for the points of the parameter space with $\Omega_{DM}=0.12$ (including both allowed and excluded points from direct detection limits). We can see that for $m_{DM} > \mu$ contours of constant $\Omega_{DM}$ have an approximately constant $c_Z$. This confirms the discussion based on analytics in section \ref{sec:totalsigmav}. Indeed, using (\ref{eq:sigmatotapprox}) with the parameter choices here, we find that for $100 \ \text{GeV} \le \mu \le 300 \ \text{GeV}$, $c_Z$ ranges from $0.005 \lsim c_Z \lsim 0.008$. These values are clearly illustrated in fig.~\ref{plot:plotSDOmega}. 

We conclude that $c_Z$ (and consequently $\sigma^{SD}$) is basically fixed by the relic density constraint. Requiring $\chi$ to be all the dark matter leads to a nearly unique prediction for the SD cross section! Fortunately, as shown in fig.~\ref{plot:plotSDOmega}, these values of $c_Z$ are still allowed by the current direct detection experiments, IceCube in particular.\footnote{It is quite crucial that our DM annihilates almost exclusively to  $t \bar{t}$ and Higgsinos. The IceCube bound on $t \bar{t}$ is by a factor of a few weaker than the $W^+W^-$ cross section, and it saves the model from being already ruled out (stable Higgsinos would not contribute to DM detection in IceCube). } With factor of 10-100 improvements in cross section expected from Xenon1T \cite{Aprile:2015uzo} and LZ \cite{Akerib:2015cja}, the next generation of DM direct detection experiments will be sensitive to essentially the entire parameter space of this model (assuming $\chi$ is a thermal relic and is all the DM). A discovery might be right around the corner!

\section{Outlook}\label{sec:Outlook}
In this section we briefly discuss the UV behavior of the model (in particular the Landau poles) and the potential sensitivity from LHC Run II. Finally we conclude with some thoughts on future directions.

\subsection{UV considerations}\label{sec:Landau Poles}
So far we have been exploring our model at the EW scale and have identified the interesting parts of the parameter space around $k_u \sim k_d \sim 1.2-1.6$. Here we want to examine the UV consequences of such large Yukawa couplings and comment on possible solutions to the Landau pole problem.

\begin{figure}
\centering
\includegraphics[scale=0.7]{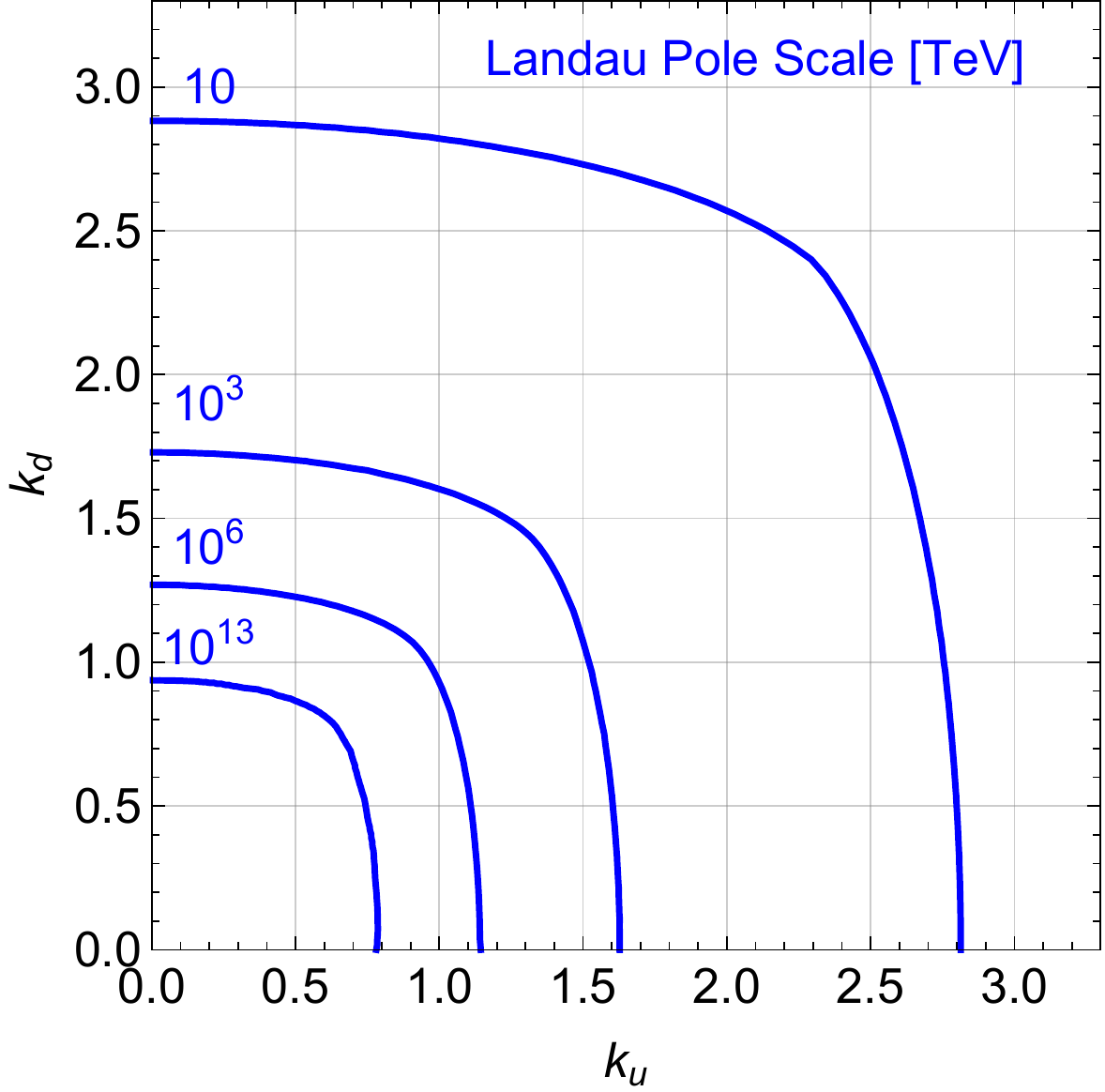}
\caption{Scale of Landau poles with one-loop RGE's in terms of $k_d, k_u$ at the IR scale. We are assuming for each point on the plot that $k_u$ and $k_d$ are given at $\Lambda_{IR} = 1 \text{ TeV}$.}
\label{plot:landau_poles}
\end{figure}

Let's focus on the most important couplings, $(g_1, g_2, g_3, y_t, k_u, k_d)$ and neglect the effect of the other couplings in finding the scale of Landau poles. Starting from one loop beta functions above the scale of the new fields (including spectator color triplets for unification) we have
\bea
\beta_{g_i} &= \frac{b_i}{16 \pi^2} g_i^3&(b_1, b_2, b_3) = (\frac{36}{5}, 2, -2)\\
\beta_{k_u} &= \frac{k_u}{16 \pi^2}(2 k_d^2 + 4 k_u^2 + 3 y_t^2 - \frac{3}{5} g_1^2 - 3 g_2^2)\\
\beta_{k_d}	&= \frac{k_d}{16 \pi^2}(4 k_d^2 + 2 k_u^2 - \frac{3}{5} g_1^2 - 3 g_2^2)\\
\beta_{y_t}	&= \frac{y_t}{16 \pi^2}(6 y_t^2 + k_u^2 - \frac{16g_3^2}{3} - \frac{13}{15} g_1^2 - 3 g_2^2)
\eea
Solving the RGE's of our model numerically, we can find the lowest scale at which one of the couplings hits its Landau pole. In fig.~\ref{plot:landau_poles} we show this scale as a function of the Yukawa couplings at $1$ TeV.

Note that for $k_u\lesssim 2$ (as we have considered in this work), the Landau poles are above $100$ TeV. Now we might ask: how can we understand physics above the Landau pole scale, or how can we postpone it to higher energies e.g. the GUT scale? One idea is to use non-Abelian gauge interactions for the new sector to reduce the beta functions of the Yukawa couplings: if we include multiple copies of $S$, $L$, $\bar{L}$ and charge them under a non-Abelian gauge group, the corresponding gauge coupling appears with negative sign in the beta function of $k_u, k_d$ (see e.g.\ \cite{Nakai:2015swg} for a recent implementation of this idea). As $S$ is Majorana, we need $S$ to be in a real representation of the new gauge group. A simple example is when the gauge group is $SO(N)$ and $S$ is in the fundamental representation. Another possibility might be to match our model to the magnetic side of a Seiberg duality and interpret physics above the scale of the Landau pole by the electric theory. It will be interesting to explore these ideas further in the future.

\subsection{LHC Phenomenology}\label{sec:LHC Phenomenology}
In addition to direct detection experiments, DM models are also probed by the LHC. 
In principle, monojet+MET \cite{Khachatryan:2014rra, Aad:2015zva} and monophoton+MET \cite{Chatrchyan:2012tea, Aad:2012fw}  searches for direct DM production could be sensitive to our model. Since quarks and gluons only talk  to $\chi$ through $s$-channel diagrams involving $Z$'s and Higgses, these searches constrain the same $c_Z$ and $c_h$ couplings as direct detection. However, these constraints   are weaker by several orders of magnitude than those from direct detection under the assumption that our DM candidate $\chi$ is all of the relic density, for the mass range we consider. See e.g.\ \cite{Zhou:2013fla, Buchmueller:2014yoa} for a recent discussion in terms of simplified DM models.

\begin{figure}[t]
\centering
\subfigure{\label{fig:mono-h}\includegraphics[width= 0.7 \textwidth]{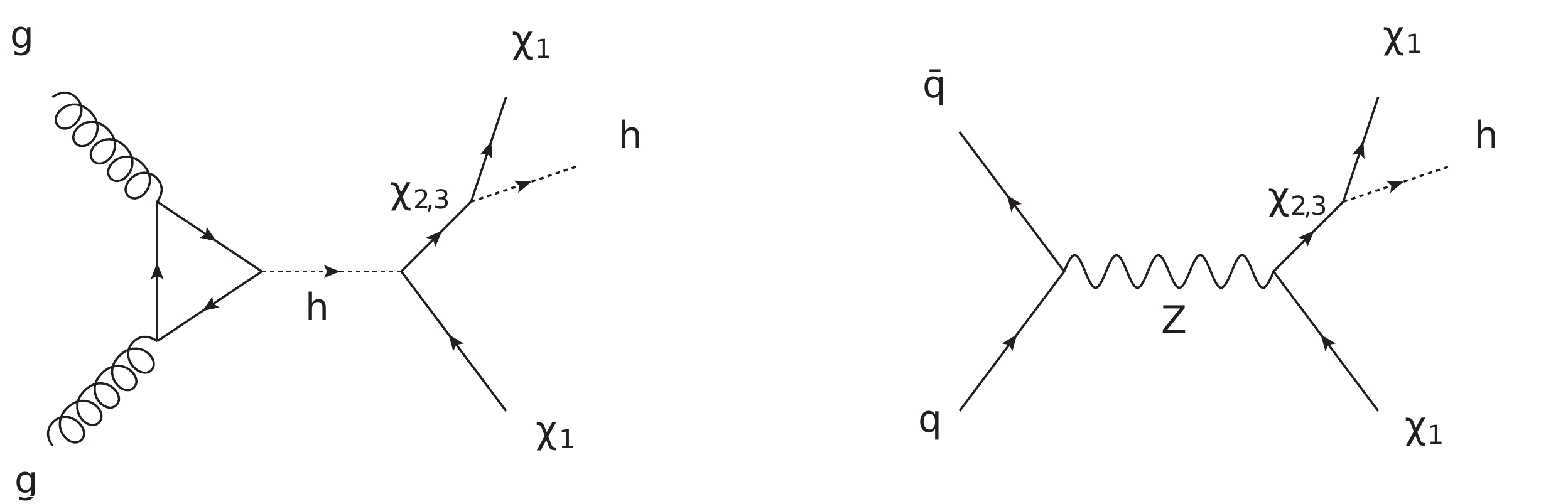}}
\subfigure{\label{fig:mono-W}\includegraphics[width=0.35 \textwidth]{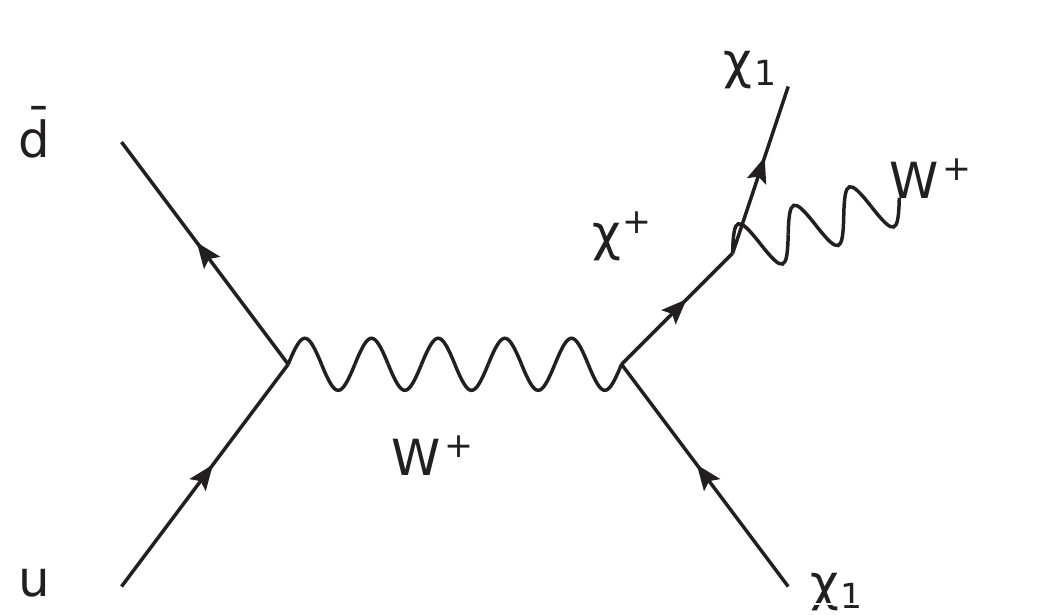}}
\caption{Diagrams contributing to mono-Higgs/Z/W+MET in our model.}
{\label{fig:mono-hZW}}
\end{figure}

Instead, let us briefly consider mono($h$, $W$, $Z$)+MET. This can occur in our model through production of $\chi_1\chi_{2,3}^0$ and $\chi_1\chi^\pm$ and subsequent decay of the (mostly-doublet) $\chi_{2,3}^0$ and $\chi^\pm$. A full treatment including estimation of SM backgrounds, detector acceptances, etc.\ is beyond the scope of this work. Here we will just present the raw production cross sections in our model. 

\begin{figure}[t]
\centering
\includegraphics[width=1\textwidth]{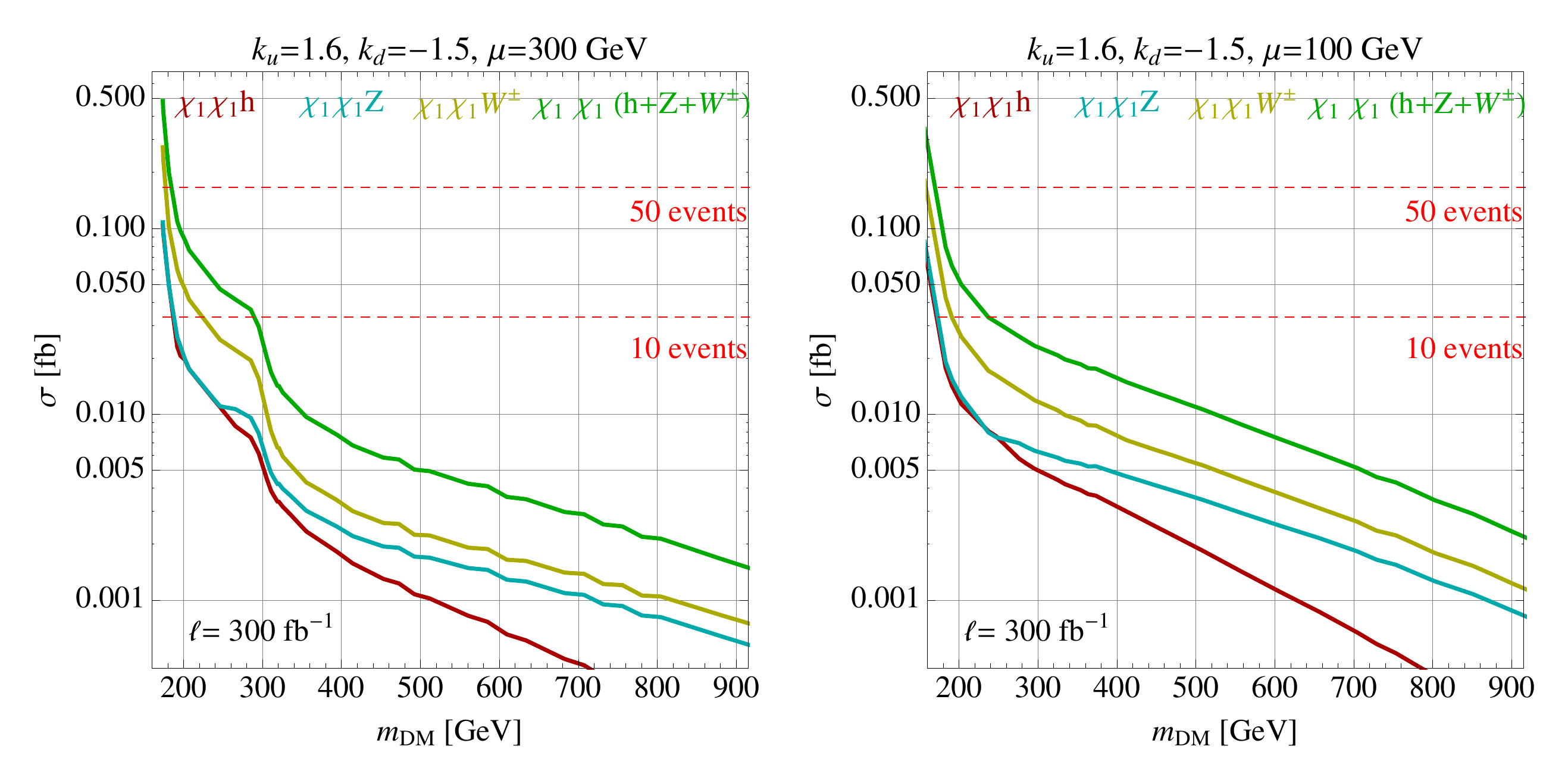}
\caption{\small{Final state cross sections for LHC13 with $\mu = 300$ GeV (left) and $\mu = 100$ GeV (right), $\tan\beta = 10$ and all superpartners decoupled. We show in green the total cross section (adding $h$, $Z$ and $W^\pm$ channels). For the $W$ cross section we are showing the sum of $W^+$ and $W^-$ final states.}}
\label{fig:LHC13}
\end{figure}

Diagrams contributing to mono-Higgs/$W$+MET are shown in fig.~\ref{fig:mono-hZW} (mono-$Z$+MET is the same as mono-Higgs with the final state Higgs replaced by $Z$). Note that we have included the one-loop gluon fusion diagram.\footnote{We thank Matt Reece for bringing this to our attention.}  Because of the large, $\CO(1)$ Yukawas $k_u$, $k_d$ in this model, this contribution can be as much as 60\% of the total $\chi_1\chi_{2,3}^0$ cross section. 
We calculated the gluon fusion contribution analytically, and the tree level contributions both analytically and with MadGraph5 \cite{Alwall:2014hca} using the model file generated by SARAH   and the spectrum files generated by SPheno. More details on the analytics are given in appendix \ref{appendix:LHC analytics}. In both cases, we used the NNPDF2.3 \cite{Ball:2012cx} PDF set. 
Fig.~\ref{fig:LHC13} shows the sum of tree level contributions and gluon fusion along the $\Omega_{DM}=0.12$ contour. We see that LHC13 will ultimately be able to probe the small mass region. Of course, if $\chi$ is all of the dark matter, then direct detection experiments will discover the model first. In that case, the LHC will only be useful as a post-discovery confirmation of the model. However, since the LHC is producing $\chi$ directly, it does not depend on the relic density. Therefore if our dark sector is only one component of $\Omega_{DM}$, the direct detection limits could be greatly relaxed while the LHC would remain sensitive.

\subsection{Future directions}

The work presented in this paper is a simple realization of a general idea: economically extending the MSSM with a single sector that provides both thermal WIMP dark matter and the 125 GeV Higgs mass. Here we took this sector to be a singlet and a pair of doublets, but one could easily imagine many other possibilities. For instance, very popular ideas for lifting the Higgs mass include the NMSSM (see e.g.\ \cite{Ellwanger:2009dp} for a review and original references) and non-decoupling D-terms \cite{Batra:2003nj,Maloney:2004rc}. While dark matter in the NMSSM is a well-studied topic, it would be very interesting to try to connect non-decoupling D-terms to dark matter. 

Even within the context of our specific singlet-doublet model, there are many interesting open questions. In this work we made some simplifying assumptions in our analysis, and it would be interesting to explore the consequences of relaxing these assumptions. 
 For example, we took all model parameters to be real, but in general there is one physical CP-violating phase. The effect of this phase on direct detection and annihilation cross sections can qualitatively change the model's behavior. Furthermore, we took all the soft mass-squareds to be positive to increase the Higgs mass. One might wonder how the phenomenology of the model would change if one of the soft masses is negative and the DM is a scalar instead of a fermion. We also assumed negligible $A$-terms in the dark sector. By analogy to stops, having substantial $A$-terms can help in raising the Higgs mass, see e.g.\ \cite{Martin:2009bg}. This could allow for smaller $k_u$, $k_d$ and open up more of the parameter space. 
 Additionally, we focused on dark matter above $\sim 100$~GeV. It could be interesting to study the phenomenology of the model for lighter dark matter masses. In particular the annihilations through the Higgs and $Z$ resonances could be large enough while still having suppressed direct detection signals. Finally, one could relax the assumption that $\chi$ is thermal and is all of the DM, and consider non-thermal relics or multi-component DM scenarios. All of these directions will become especially well-motivated if nothing is discovered at the next round of direct detection experiments, as discussed in section \ref{sec:results}.

There are also many interesting model-building directions in the UV. For example, enlarging the dark sector to accommodate a non-Abelian gauge symmetry could have potentially interesting consequences. As noted in section \ref{sec:Landau Poles}, this may help postpone the Landau pole of the Yukawa couplings, and the new gauge interactions could play an important role in the dynamics of the dark sector. Additionally we have two supersymmetric masses $M_L$ and $M_S$ at the electroweak scale. Perhaps the same dynamics that generates $\mu$ in the MSSM is responsible for generating these masses as well.


\section*{Acknowledgements:}
We thank Kyu Jung Bae, Matt Buckley, John Paul Chou, Eva Halkiadakis, Ryuichiro Kitano, Stephen Martin, Natsumi Nagata, Matt Reece, Alessandro Strumia and Scott Thomas for helpful discussions. We are especially grateful to Florian Staub for his indispensable help setting up the model in SARAH. We are also grateful to Natsumi Nagata and Florian Staub for comments on the manuscript.  This work was supported by DOE grant DE-SC0013678.

\appendix

\section{Connecting model parameters to DD cross sections}\label{appendix:DD_to_ch/cZ}

In this appendix, we will review how to relate the SI and SD DM-nucleon cross sections to the couplings $\xi_q^{SI}$ and $\xi_q^{SD}$ appearing in the effective Lagrangian (\ref{eq:dm quark lagrangian}). To check our results we verify that by calculating SI and SD cross sections analytically, we get the same result as the one we get from micrOMEGAs.

Following  \cite{Jungman:1995df}, the SI and SD cross sections are
\bea\label{eq:SISDxsec}
\s^{SI}_{p,n}&=\frac{ x m_r^2}{\pi} f_{p,n}^2,\qquad \s^{SD}_{p,n}&=\frac{ 3 x m_r^2}{\pi}a_{p,n}^2
\eea
where $x=4$ for Majorana ($x=1$ for Dirac) fermions, $m_r$ is the reduced mass of the DM-nucleon system, and $a^{SD}_{p,n}, f^{SI}_{p,n}$ are the effective DM-nucleon couplings: 
\bea\label{eq:SISDformfactor}
f_{p,n}&=\sum_{q=u,d,s} \xi_q^{SI} f^{p,n}_q \frac{m_{p,n}}{m_q}+\frac{2}{27} (1-\sum_{q=u,d,s} f^{p,n}_{q}) \sum_{q=c,b,t} \xi_q^{SI} \frac{m_{p,n}}{m_q}\\
a_{p,n}&=\sum_{q=u,d,s} \xi_q^{SD} \,\Delta_q^{p,n},
\eea
Here $m_q$ is the quark mass, $f^{p,n}_q$, and $\Delta_q$ are hadronic parameters calculated for example by lattice simulations in QCD. We use the values in tab.~\ref{tab:SD form factors} according to \cite{Belanger:2013oya}. The difference between SI cross sections for proton and neutron is negligible as the main contribution comes from $f_s$ which is the same in both cases. 

\begin{table}[t]
$$
\begin{array}{|c|c|c|c|c|c|c|}\hline
 & \Delta_u & \Delta_d & \Delta_s & f_u & f_d & f_s \\\hline
 p& 0.842 & -0.427 & -0.085 & 0.0153 & 0.0191 & 0.0447 \\ \hline
 n& -0.427 & 0.842 & -0.085 & 0.011 & 0.0273 & 0.0447 \\ \hline
\end{array}
$$
\caption{Nucleon quark form factors.}
\label{tab:SD form factors}
\end{table}

\begin{figure}
\centering
\includegraphics[scale=0.62]{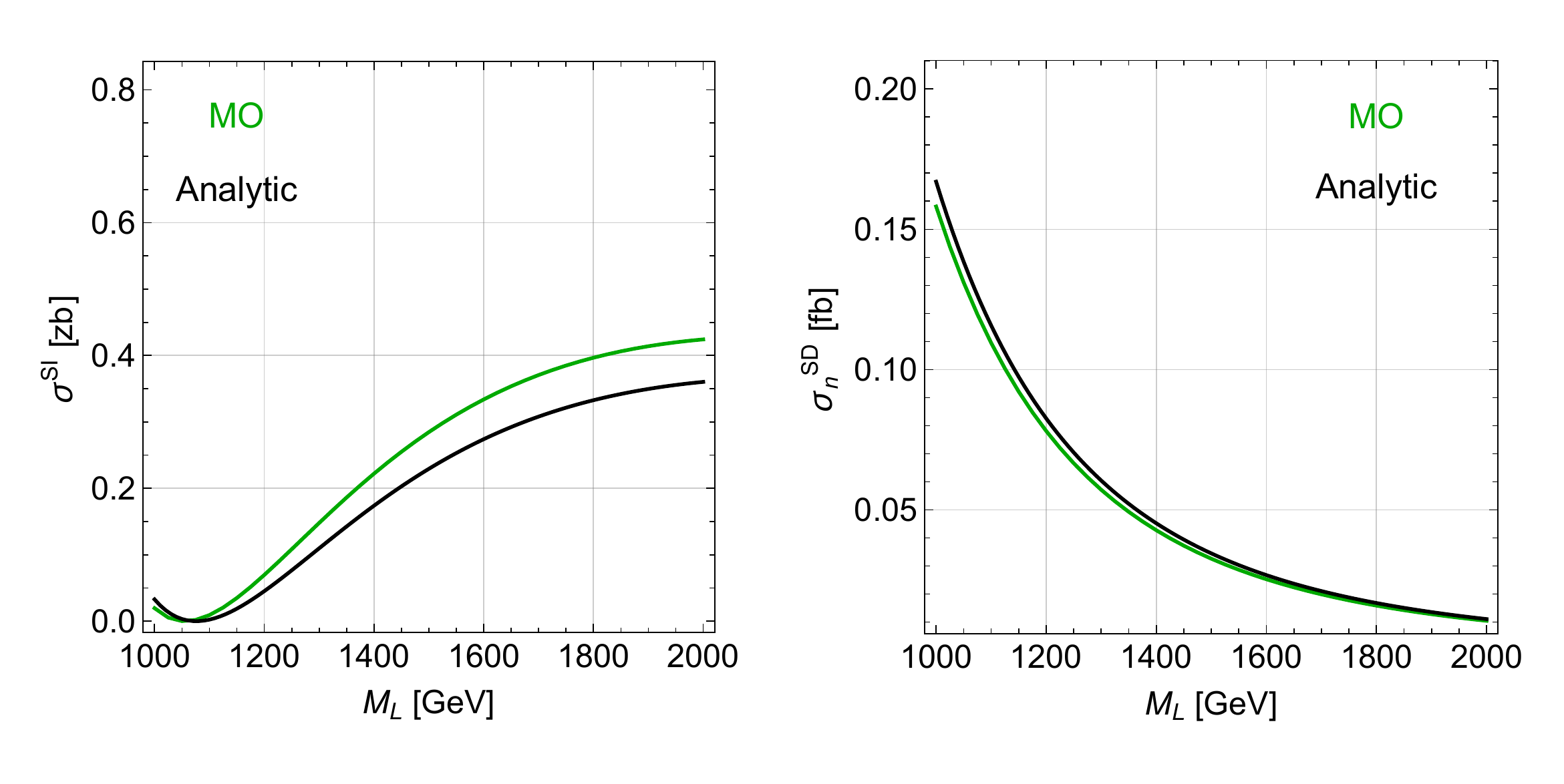}
\caption{\small{Comparing (\ref{eq:SISDxsec}) with micrOMEGAs 4.1.8 for $M_S=200$ GeV, $k_u=1.6$, $k_d = -1.5$.}}
\label{fig:MO DD comparison}
\end{figure}

In fig.~\ref{fig:MO DD comparison}, we compare our analytic cross sections to micrOMEGAs. We see that the agreement is excellent.  

\section{Validating SPheno one loop Higgs mass}\label{appendix:Validating_Higgs_mass}

In this appendix we validate the contributions to the Higgs mass  from the dark sector as calculated by SPheno against the analytic one-loop calculation through the Coleman-Weinberg (CW) potential. We consider the simplified one-loop CW result from (\ref{eq:deltamhsq}), where we ignored $g_{1,2}$ and $\mu$ and took the $\tan\beta\to\infty$ limit. We will show that both one-loop and two-loop results from SPheno match quite well with our analytical result.  As SPheno outputs the total Higgs mass and not just the contributions from the dark sector, we extract these contributions as follows:
\bea
\delta m_h^2 = m_h^2 - m_h^2|_{MSSM}
\eea
where $m_h^2|_{MSSM}$ is the contribution to $m_h^2$ from the MSSM with superpartners at 1 TeV.

\begin{figure}
\centering
\includegraphics[scale=0.5]{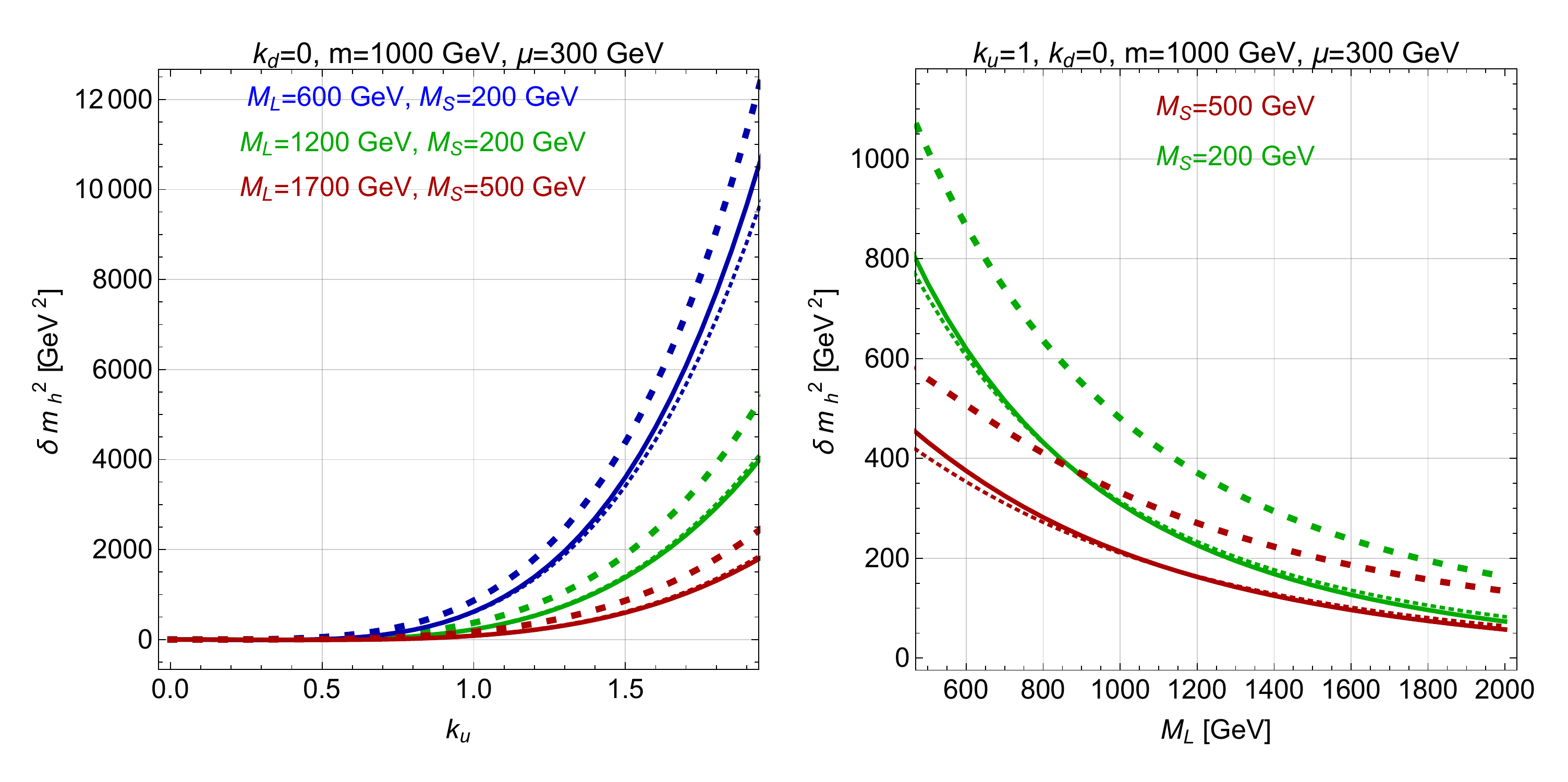}
\caption{Thick dashed: analytical one-loop result (\ref{eq:deltamhsq}). Dotted: SPheno one-loop. Solid: SPheno two-loop. }
\label{fig: SPheno_validation}
\end{figure}

Since the Higgs mass depends primarily on $k_u, M_L, M_S$, we will demonstrate here that SPheno and our CW calculation agree well as these parameters are varied. 
From (\ref{eq:deltamhsq}) we expect $\delta m_h^2 \sim k_u^4$. As we can see in fig.~\ref{fig: SPheno_validation} (left), SPheno confirms this behavior. 
After fixing the $k_u$ dependence, we need to check that our analytical equations and SPheno match as we change $M_L$ and $M_S$. That is shown in fig.~\ref{fig: SPheno_validation} (right) for two values of $M_S$ as we scan over $M_L$.

\section{LHC cross section analytics}\label{appendix:LHC analytics}

Here we will provide some analytic details for the calculation of the $pp\to \chi_1\chi_{2,3}$ and $pp\to\chi_1\chi^\pm$ LHC cross sections useful for section \ref{sec:LHC Phenomenology}. The former receives contributions from both tree-level $Z$'s with $q\bar q$ initial state, as well as one-loop gluon fusion. The latter comes from tree-level $W^\pm$'s with $q \bar q'$ initial state. 

The tree-level, quark-initiated cross sections are given by:
\bea
\sigma(q\bar{q} \rightarrow \chi_1\chi_{i}) &= \frac{g_2^4|\vec{p}_{f}|}{144 \pi c_W^4 S^{3/2}}\left(c_{Zq_Lq_L}^2+c_{Zq_Rq_R}^2\right) \times\\
&((\Re{R_{i}})^2f_Z(S,m_1,m_{i})+(\Im{R_{i}})^2f_Z(S,m_1,-m_{i}))\\
\sigma(q\bar{q}'\rightarrow \chi_1\chi^+) &= \frac{g_2^4 |\vec{p_f}|}{576 \pi S^{3/2}} \left( f_W(S,m_1,m_+) |R_+|^2 + f_W(S,m_1,-m_+)|R_-|^2\right)
\eea
where
\bea
R_{i}&=(U_{1,2}^*U_{i,2}-U_{1,3}^*U_{i,3})\\
R_\pm &=U_{1,2}\pm U_{1,3}^*\\
f_A(S,x,y)&=\frac{(S-(x+y)^2)(S+\frac{(x-y)^2}{2})}{(S-m_A^2)^2}
\eea
and
\beq
c_{Z \bar{u}_Lu_L} = \frac{1}{2}-\frac{2}{3} s_W^2,\quad c_{Z \bar{d}_Ld_L} =  -\frac{1}{2}+{1\over3}s_W^2,\quad
c_{Z \bar{u}_Ru_R} =  \frac{2}{3} s_W^2,\quad c_{Z \bar{d}_Rd_R} =  -{1\over3} s_W^2
\eeq

The parton level gluon fusion cross section (as can be calculated e.g.\ using \cite{Gunion:1989we}) is
\bea
\sigma(g g \rightarrow \chi_1\chi_i) &= \frac{|\vec{p_f}| m_t^2}{128 \pi S^{5/2}}\left|{\lambda_t \alpha_3 c_{\chi_1\chi_i h}F(S/m_t^2)\over 4\pi}\right|^2     \, g(S,m_1,m_i)
\eea
where
\bea
g(S,m_1,m_2)&={(1-\frac{(m_1+m_2)^2}{S})\over (1-\frac{m_h^2}{S})^2}\\
F(x) &= 2\sqrt{2}\left(1+\left(1-{4\over x}\right)\left[\sin^{-1}\sqrt{x\over 4}\right]^2\right) .
\eea
and $c_{\chi_1\chi_i h}$ is the coupling between Higgs and $\chi_1 \chi_i$ ($i = 2, 3$) defined in the same way as $c_h$ in (\ref{eq:chcZL}):
\bea
\mathcal L &\supset c_{\chi_1\chi_i h} h \bar\psi_{\chi_1}\psi_{\chi_i}\\
c_{\chi_1\chi_i h} &= \frac{1}{\sqrt{2}}\left(\hat k_u (U^*_{1,1} U^*_{i,2}+U^*_{1,2} U^*_{i,1}) + \hat k_d  (U^*_{1,1} U^*_{i,3}+U^*_{1,3} U^*_{i,1}) \right),
\eea

\bibliography{DM_Higgs_bib}

\end{document}